\pdfoutput=1
\documentclass[aps,prx,superscriptaddress,twocolumn]{revtex4-1}

\makeatother

\usepackage[dvipsnames]{xcolor}
\usepackage{amsmath}
\usepackage{amssymb}
\usepackage{graphicx}
\usepackage{bm}
\usepackage{float}
\usepackage{physics}

\usepackage{lipsum}

\usepackage[colorlinks=true,bookmarks=false,citecolor=blue,linkcolor=red,hyperfootnotes=true,urlcolor=blue]{hyperref}

\usepackage{user} 

\begin{document}

\title{Machine learning of Kondo physics using variational autoencoders and symbolic regression}

\author{Cole Miles}
\email[]{cmm572@cornell.edu}
\affiliation{Department of Physics, Cornell University, Ithaca, New York 14853, USA}

\author{Matthew R. Carbone}
\affiliation{Computational Science Initiative, Brookhaven National Laboratory, Upton, New York 11973, USA}

\author{Erica J. Sturm}
\affiliation{Condensed Matter Physics and Materials Science Division, Brookhaven National Laboratory, Upton, New York 11973, USA}

\author{Deyu Lu}
\affiliation{Center for Functional Nanomaterials, Brookhaven National Laboratory, Upton, New York 11973, USA}

\author{Andreas Weichselbaum}
\affiliation{Condensed Matter Physics and Materials Science Division, Brookhaven National Laboratory, Upton, New York 11973, USA}

\author{Kipton Barros}
\affiliation{Theoretical Division and CNLS, Los Alamos National Laboratory, Los Alamos, New Mexico 87545, USA}

\author{Robert M. Konik}
\affiliation{Condensed Matter Physics and Materials Science Division, Brookhaven National Laboratory, Upton, New York 11973, USA}

\date{\today}

\begin{abstract}
We employ variational autoencoders to extract physical insight from a dataset of one-particle Anderson impurity model spectral functions. Autoencoders are trained to find a low-dimensional, latent space representation that faithfully characterizes each element of the training set, as measured by a reconstruction error. Variational autoencoders, a probabilistic generalization of standard autoencoders, further condition the learned latent space to promote highly interpretable features. In our study, we find that the learned latent variables strongly correlate with well known, but nontrivial, parameters that characterize emergent behaviors in the Anderson impurity model. In particular, one latent variable correlates with particle-hole asymmetry, while another is in near one-to-one correspondence with the Kondo temperature, a dynamically generated low-energy scale in the impurity model. Using symbolic regression, we model this variable as a function of the known bare physical input parameters and ``rediscover'' the non-perturbative formula for the Kondo temperature. The machine learning pipeline we develop suggests a general purpose approach which opens opportunities to discover new domain knowledge in other physical systems.
\end{abstract}

\maketitle

\section{Introduction}

Experimental spectra such as those obtained by x-ray absorption, resonant inelastic x-ray scattering, optics, and angle resolved photoemission provide detailed information about the system response for strongly correlated materials~\cite{Damascelli03,Valla99,Dean16}.
Spectroscopy techniques typically involve the absorption and emission of energy that induce transitions between states of the material. These processes are generally expensive/time-consuming to measure or simulate, and are also difficult to interpret, as they involve complicated many-body interactions.

A particularly important quantity of interest is the one-particle
Green's function, $G(\omega),$ which characterizes the many-body system's response to
the injection or removal of an electron~\cite{mahan2013many}. The spectral function,
$A(\omega) = -\frac{1}{\pi}\Im G(\omega),$ completely encodes $G(\omega)$ through Kramers-Kronig relations and is generally experimentally measurable.
Its peaks and line shapes encode non-trivial many-body features such as renormalized
energy scales, the many-body eigenspectrum, and quasiparticle lifetimes.
However, interpretation of spectral data is often a significant and system-dependent challenge requiring the combination of experimental
spectroscopy, educated guesses, and informed theoretical analysis.
In this paper, we develop a data-driven representation learning workflow for generating physical insight in the form of analytic expressions from a collection of physical data, and demonstrate its power in this context of spectral interpretation.

The goal of representation learning techniques is to automatically extract a low-dimensional set of features that describe the trends and variations of a collection of high-dimensional raw data~\cite{bengio_representation_2013}.
Unlike supervised learning, each datapoint remains unlabeled, and the goal is to
learn the underlying structure of a dataset as opposed to a 1-to-1 mapping. While there exist many viable techniques for representation learning, we focus here on the recently prominent approach of training autoencoders \cite{Tschannen2018ArXiv181205069CsStat}.
These are feed-forward neural networks trained to approximate the identity function $f(\vecx) \approx \vecx$, where the network has a restricted functional form that involves a low-dimensional ``bottleneck'' (latent space). The intention is that the latent space should distill the information most necessary for accurate reconstruction of the input. In this work, we use a stochastic variant called the \textit{variational autoencoder} (VAE)~\cite{kingma_auto-encoding_2014, doersch_tutorial_2021} which has been empirically found to improve interpretability of the latent space components \cite{higgins2016beta}.

Autoencoder-based unsupervised learning has begun to see wide use in the physical sciences in various applications such as generating realistic data following a learned distribution~\cite{Martinez-Palomera2020ArXiv200507773Astro-Ph,Otten2021NatCommun}, collider event anomaly detection~\cite{Farina2020Phys.Rev.D, Cerri2019J.HighEnerg.Phys.}, phase identification~\cite{Wetzel2017Phys.Rev.E, Kottmann2020Phys.Rev.Lett.}, and as a tool in inverse optimization problems~\cite{Samarakoon2020NatCommun}. However, the prospect of using this technique for the \textit{discovery} of interpretable physical features has been explored only very recently~\cite{routh_latent_2021, Lu2020Phys.Rev.X, Kalinin2021Sci.Adv.}. In these works, VAEs were found to produce remarkably interpretable low-dimensional features characterizing datasets \cite{routh_latent_2021, Kalinin2021Sci.Adv.}, even near-perfectly reproducing known physical parameters in controlled settings \cite{Lu2020Phys.Rev.X}.
But it is not clear how to best train these models or connect them to unknown physics without significant prior knowledge. We here explore these questions and demonstrate a complete pipeline to train interpretable VAEs, and additionally propose a novel application of symbolic regression~\cite{schmidt_distilling_2009,bongard_automated_2007} to discover analytic expressions describing the captured features in terms of a set of known physical quantities. 

The present study focuses on a dataset of
spectral response functions from the famous single impurity Anderson model (SIAM)~\cite{Anderson61}. The SIAM model
describes a single quantum impurity (or quantum dot) embedded in a non-interacting bath of fermions. Despite this model's simplicity, it is a key tool in the
examination of strongly correlated low-energy phenomena such as
Kondo screening and the Kondo resonance.
We have selected this system as it is well-understood and numerically inexpensive to evaluate non-perturbatively. Thus the creation of a large, varied dataset for
machine learning (ML) purposes is achievable. Moreover, the dynamically generated low-energy regime of this model is well-known to be non-analytically related to the bare parameters, presenting a non-trivial set of features for the VAE to unravel.

The outline of the paper is as follows. We first introduce the model
and review VAEs in the context of spectral function reconstruction. Then, we demonstrate that training VAEs with varying regularization strengths reveals a ``critical'' number of dimensions needed to characterize a dataset. We find that each dimension of the latent space corresponds to a key physical descriptor for the set of SIAM spectral functions. The automatically discovered descriptors include (a) the Kondo temperature~\cite{Kondo1964}, (b) a measure of particle-hole asymmetry, and (c) the presence of competing energy scales. Finally, we propose and demonstrate the use of symbolic regression to extract analytic expressions for these emergent physical descriptors.

\section{Methodology}

\subsection{Overview of physical system}

We examine the spectral functions of the 
prototypical SIAM with Hamiltonian
\begin{subequations}
\label{Eq-SIAM}
\begin{equation} 
    \hat{H} = \hat{H}_\mathrm{imp}
    + \hat{H}_\mathrm{bath}
    + \hat{V}
\end{equation} 
where
\begin{eqnarray}
    \hat{H}_\mathrm{imp} &=& \sum_\sigma \epsilon_{d\sigma} \hat{n}_\sigma + U\hat{n}_{\uparrow}\hat{n}_{\downarrow},
    \quad \hat{n}_\sigma = d_\sigma^\dagger d_\sigma,
\\
    \hat{H}_\mathrm{bath} &=& \sum_\sigma \int_{-D}^D \dif \epsilon \: \epsilon \,\hat{c}_{\epsilon\sigma}^\dagger\hat{c}_{\epsilon\sigma},
\\
    \hat{V} &=& \sum_\sigma \int_{-D}^D \dif\epsilon \: \sqrt{\tfrac{\Gamma(\epsilon)}{\pi}}
    \left(\hat{d}_\sigma^\dagger\hat{c}_{\epsilon\sigma} + \mathrm{H.c.}\right).
\end{eqnarray}
\end{subequations}
The impurity (imp) constitutes a particle with spin $\sigma=\pm 1 (\equiv \uparrow\downarrow)$ at energy
level $\epsilon_{d\sigma} = \epsilon_d{-}\frac{\sigma}{2}B$ relative to the
Fermi surface in the presence of an external magnetic field of strength $B.$
Double occupation is penalized by the Coulomb repulsion energy $U.$
The impurity couples to the bath via $\hat{V}$ characterized by the
hybridization function $\Gamma(\epsilon) = \pi\rho_\epsilon V^2_\epsilon$, where
$\rho_\epsilon$ is the density of states and $V_\epsilon$ represents the hopping
elements between the impurity and bath level at energy $\epsilon$ relative to
the Fermi energy, with the normalized bath levels obeying
$\{\hat{c}_{\epsilon\sigma}, 
\hat{c}_{\epsilon'\sigma'}^\dagger\} =
\delta(\epsilon - \epsilon')\,\delta_{\sigma\sigma'}$.
The fermionic bath is described by $\hat{H}_\mathrm{bath}.$
In this work we examine the SIAM spectral functions 
for constant hybridization functions within the bandwidth, 
$\Gamma(\epsilon) = \Gamma\vartheta(D-\abs{\epsilon})$,
i.e. the so-called ``box'' distribution, where $\vartheta$ is the
Heaviside step function. Hence the model Hamiltonian
\eqref{Eq-SIAM} is particle-hole
symmetric for $\epsilon_d = -U/2$.
Unless otherwise noted, all energies are 
in units of the half-bandwidth $D=1$. We also set $\hbar=k_\mathrm{B}=1$.

The spectral function
$A_\sigma(\omega) = -\tfrac{1}{\pi}\, {\Im} G^R_\sigma(\omega)$ represents the  impurity's local density of states,
as derived from the time-dependent retarded fermionic Green's
function $G^R_\sigma(t) {=} -\iu
\vartheta(t) \langle \{ \hat{d}_\sigma(t),
\hat{d}_\sigma^\dagger \} \rangle_T$ after Fourier
transform from time $t$ to frequency $\omega$,
with $\langle \cdot \rangle_T$ denoting the thermal average.
At finite $B$,
we pick an arbitrary but fixed spin orientation
$\sigma$ throughout.
The features of the spectral data are well-known
analytically:
At $B=0$, the spectral data shows peaks
around $\omega=\epsilon_d$ and $\omega=\epsilon_d+U$
of width $\Gamma$ (the so-called Hubbard side peaks)
which are trivially related to bare energies of the
Hamiltonian,
as well as a peak pinned around $\omega=0$ which derives
from low-energy particle-hole excitations mediated by
the Kondo interaction. The width of the latter peak defines the dynamically
generated low-energy Kondo scale $T_K$, where for the Kondo regime
typically $T_K \ll \Gamma < U$. This peak gets suppressed
by temperature $T>T_K$, and becomes strongly asymmetric
in the spin-resolved case
when applying a magnetic field $|B|>T_K$.

Each spectral function $A(\omega)$ is generated
by the numerical renormalization group (NRG)~\cite{Wilson75,Bulla08,Wb07,Weichselbaum2012},
and is controlled by five independent physical parameters: $U,$ $\Gamma,$
$\epsilon_d,$ $B,$ and $T$ (the temperature). Values for each of the parameters
are chosen randomly according to the Anderson set procedures presented
in Ref.~\cite{Sturm21}. To normalize the spectra to roughly comparable absolute scales, we utilize our knowledge of the Friedel-sum rule to motivate a normalization of the spectra to $\vecx \equiv \pi\Gamma A(\omega_i) \in [0, 1]$.
As a second element of human input, we choose to sample the spectra on a fixed linear-logarithmic
frequency grid $\omega_i$ to form the inputs to the VAE.
This frequency
sampling, as detailed in Fig.~\ref{fig:omega_grid} and Ref.~\cite{Sturm21}, allows variations of the dynamically-generated low-energy scale at exponentially small $\omega$ to be clearly resolved.

\begin{figure*}
\centering
\includegraphics[width=0.8\textwidth]{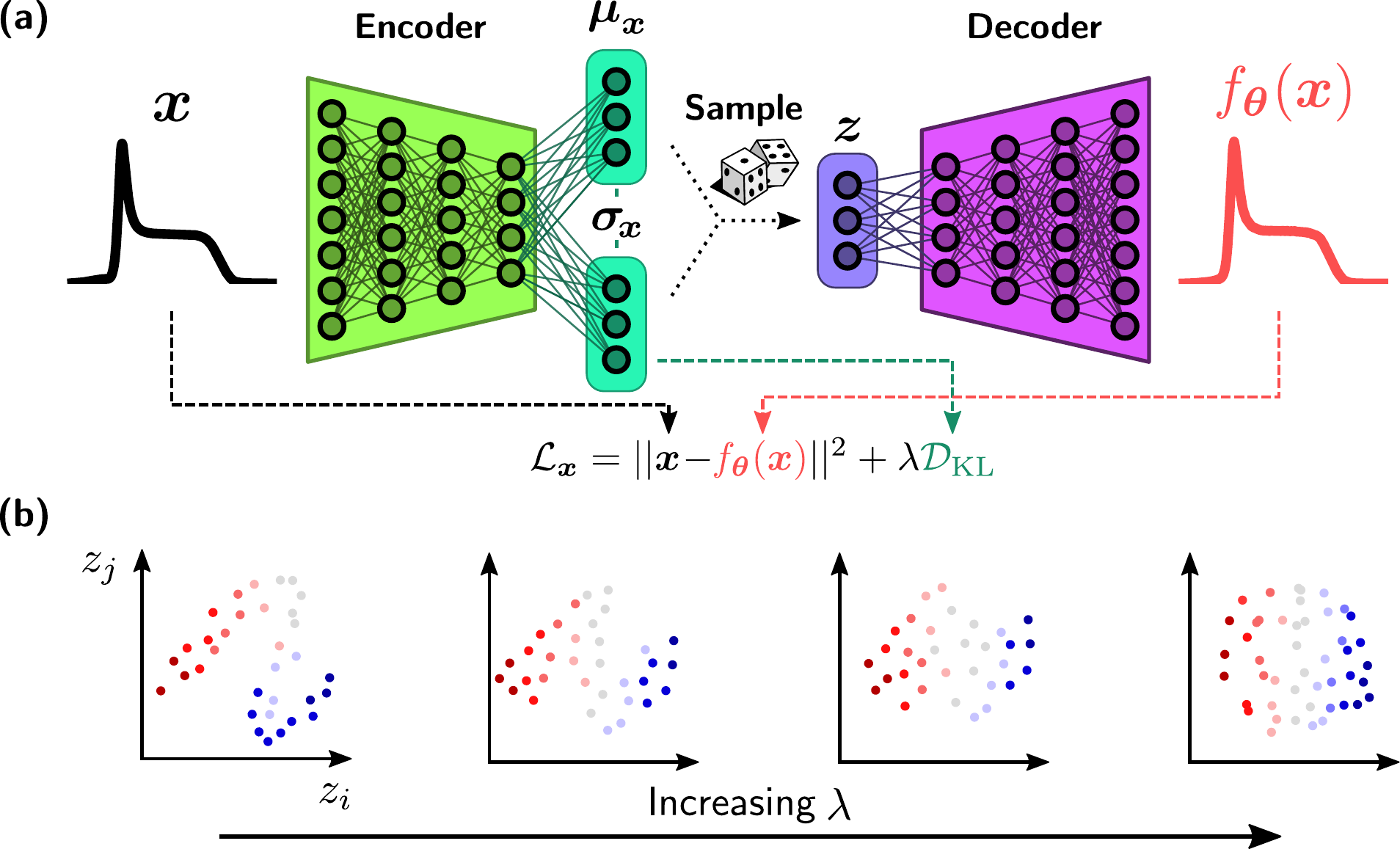}
\caption{%
    (a) The variational autoencoder architecture. Spectral functions
    $\vecx$ on a pointwise grid are fed into the encoder and compressed
    into parameters $\boldsymbol{\mu}_\vecx, \boldsymbol{\sigma}_\vecx$ of an $L$-dimensional Gaussian distribution in latent space. Latent activations $\vecz$ are then sampled from $\vecz \sim \mathcal{N}(\boldsymbol{\mu}_\vecx, \mathrm{diag}(\boldsymbol{\sigma}_\vecx^2))$ before being fed into the decoder, which reconstructs the input spectrum. The contribution to the loss from a given $\vecx$, $\mathcal{L}_\vecx$ is shown in brief, with arrows
    noting which components of pipeline contribute. (b) A schematic showing that increasing $\lambda$ structures the data distribution in latent space such that the $z_i$ become statistically independent, and aligns underlying generative factors (red/blue coloring) with the latent axes.
}
\label{fig:explain_vae}
\end{figure*}

We utilize the smallest physical energy (SPE) value,
$E_0 \equiv \mathrm{max}\left(|B|, T, T_K\right)$
to define two datasets with different effective dimensionalities in their respective physical parameter spaces.
Here $T_K$ is the Kondo
temperature known analytically 
at $B=T=0$ as \cite{Haldane78,tsvelik_weigmann} 
\begin{equation}
    T_K = \sqrt{\tfrac{U\Gamma}{2}}\exp\left\{\tfrac{\pi \epsilon_d (\epsilon_d + U)}{2 U\Gamma}\right\}.
    \label{eq:TK}
\end{equation}
The first dataset, referred to as the $T_K$-dominated
dataset, is constrained to spectral functions within the parameter space
$|B|,T < T_K / 20$ \footnote{This cutoff in $T_K$ was heuristically chosen to both minimize the effects of $B, T$ while keeping the dataset roughly the same size as our full-parameter dataset.}, with a total of about $28,000$ spectral functions present in the training set. Spectra from this dataset are thus effectively 
described by three parameters $(U,\Gamma,\epsilon_d)$. The second dataset is unconstrained, such that the SPE is equally controlled by $\lvert B\rvert, T, \text{or } T_K$, 
resulting in a total of five parameters. We refer to this dataset as
the full-parameter dataset, with a total of about $60,000$ spectral
functions. The random sampling of these two datasets from the data 
pool of Ref.~\cite{Sturm21} is done independently, so that the first dataset is not a 
subset of the second.

\subsection{Overview of variational autoencoders}

The core idea of all autoencoder-based architectures is to learn an effective compression of a dataset by learning a parameterized form of the identity function with an informational bottleneck. As shown in Fig.~\ref{fig:explain_vae}, VAEs do this using two components: the encoder and the decoder. A standard fully-connected encoder compresses the input through a series of affine transformations followed by point-wise nonlinear functions. The output of the $m$\textsuperscript{th} layer $\veca^{(m)}$ is given by
\begin{equation} \label{eq: ML layer activation structure}
    \veca^{(m)} = \phi\bigl( W^{(m)} \veca^{(m-1)} + \vecb^{(m)} \bigr),
    \quad \veca^{(0)} \equiv \vecx,
\end{equation}
where $\phi$ is a nonlinear activation function acting element-wise on its
input. All encoder layers rely upon the rectified linear unit activation
function ($\mathrm{ReLU}\left(a\right) = \max(0,a)$), except the final layer
which uses the identity function. The encoder consists of $D_\mathrm{enc}$ total hidden layers.

The weight matrices $W$ and bias vectors $\mathbf{b}$ are all free, learned parameters
which we collectively denote as $\boldsymbol{\theta}$.
In the encoding stage, the number of activations
$\veca^{(m)}$ steadily decreases, forcing the model to learn consecutively
lower-dimensional representations of the input data. We denote the final values output by the encoder as $\vecl \equiv \veca^{(D_\mathrm{enc})} =
[l_1, l_2, ..., l_{2L}]$, with $L$ an architectural
hyperparameter.

VAEs are distinguished from traditional autoencoders
in the sense that these final activations no longer represent a
single compressed point encoding the input $x$.
Instead, these activations are used to parameterize a normal distribution
with mean vector $\boldsymbol{\mu} = [l_1, l_2, ..., l_L]$
and log-variances $\boldsymbol{\sigma}^2 = [\exp(l_{L+1}), \exp(l_{L+2}), ..., \exp(l_{2L})]$. The $L$-dimensional space this distribution lives in is called \textit{latent space}, with each dimension being a latent variable.

On any given forward pass, the encoder maps $\vecx \mapsto
[\boldsymbol{\mu}_\vecx, \ln
\boldsymbol{\sigma}^2_\vecx]$, then samples from
the obtained multivariate normal distribution
$P(\vecz | \vecx) = \mathcal{N}(\boldsymbol{\mu}_\vecx, \text{diag}(\boldsymbol{\sigma}_\vecx^2))$, where $\text{diag}$ constructs a diagonal matrix with $\boldsymbol{\sigma}_\vecx^2$ along the diagonal.
The sampled $L$-component latent vector $\vecz$ is finally passed
to the decoder which generates the reconstruction
$\vecz \mapsto \tilde{\vecx}$. As is common practice, for simplicity and speed we only sample a single $\vecz$ for each input $\vecx$ per forward pass during training.
Combined with the training procedure outlined in the next section, this sampling process forces the VAE to acquire the notion of continuity and statistical independence in the learned latent space.
Thus, it is generally found that VAEs learn more ``meaningful'', statistically disentangled representations than standard autoencoders~\cite{higgins_early_2016}.

The decoder is a deterministic fully-connected network of $D_\mathrm{dec}$ hidden layers from the sampled latent variables $\vecz$ back to the original input space, generally following a reversed structure to the encoder as shown in Fig.~\ref{fig:explain_vae}. The goal is for the decoded output to be a minimally-lossy reconstruction for the original input $\vecx$. We denote the full VAE action as $f_{\boldsymbol{\theta}}(\vecx) \approx \vecx$, which is a random variable due to the sampling within latent space.
As before, all layers of the decoder except for the last use the ReLU activation function. For the last layer of the decoder we utilize $\mathrm{Softplus}(x) = \log(1 + \exp(x))$ as the activation function to enforce positivity of $\tilde{\vecx}$ in a smooth manner. In this work, all networks studied have $D_{\mathrm{enc}} = D_{\mathrm{dec}} = 5$, with hidden layer sizes manually tuned to
$[240, 160, 80, 60, 2L]$ (reversed for the decoder, except for $2L\to L$). We found $L = 10$ to be sufficiently large to saturate reconstruction performance on both datasets (see \App{SM:Training_details}).

\subsection{Loss functions and training}

The parameters $\boldsymbol{\theta}$ of the VAE are learned during training so as to minimize
the objective loss function, $\mathcal{L}(\boldsymbol{\theta}; \lambda).$ This loss
originates as a bound on the log-likelihood of the distribution defined by the VAE (see Sec.~\ref{SM:VAE_framework}), and can be written in our case as:
\begin{equation} \label{eq:loss:tot}
    \mathcal{L}(\boldsymbol{\theta}; \lambda) =
    \mathcal{L}_{\mathrm{RL}}(\boldsymbol{\theta}) + \lambda
    \mathcal{L}_\mathrm{KLD}(\boldsymbol{\theta}),
\end{equation}
where explicit dependence on the dataset has been suppressed for brevity. The two contributions are defined as
\begin{subequations}
\begin{align}
    \mathcal{L}_{\mathrm{RL}}(\boldsymbol{\theta})
    &= \frac{1}{N} \sum_\vecx \mathbb{E}_{\vecz}\left[
      \left|\vecx - f_{\boldsymbol{\theta}}(\vecx) \right|^2\right],
\label{eq:loss:L2} \\
    \mathcal{L}_\mathrm{KLD}(\boldsymbol{\theta})
    &= \frac{1}{N} \sum_{\vecx}
      \mathcal{D}_\mathrm{KL} \left[
     \mathcal{N}(\boldsymbol{\mu}_\vecx, \text{diag}(\boldsymbol{\sigma}^2_{\!\vecx})) \Vert
     \mathcal{N}(\mathbf{0}, I) \right],\label{eq:loss:KL}
\end{align}
\end{subequations}
and are referred to as the reconstruction loss
and the Kullback-Leibler divergence (KLD), respectively,
each averaged across the dataset of size $N$.

$\mathcal{L}_{\mathrm{RL}}$ measures the expected squared error between the input $\vecx$
and the reconstructed approximation $f_{\boldsymbol{\theta}}(\vecx)$, with the expectation over samples of $\vecz$ in latent space.
Meanwhile, $\mathcal{L}_\mathrm{KLD}$ measures the average KLD
$\mathcal{D}_\mathrm{KL}[\bullet\Vert\circ]$ 
between the distribution in latent space predicted by the encoder, and a Gaussian distribution with zero mean and unit variance~\cite{kingma_auto-encoding_2014}. For the case in \Eq{eq:loss:KL},
this term is known analytically as a function of $\boldsymbol{\mu}_{\vecx}$ and $\boldsymbol{\sigma}_{\vecx}$ and can be expressed as a sum over latent variables,
\begin{equation} \label{eq:KL_loss}
    \mathcal{D}
    _{\mathrm{KL}} = \sum_{i=1}^L
    \underbrace{\tfrac{1}{2}
    \bigl(\mu_{\vecx, i}^2 + \sigma^2_{\vecx, i}
    - 1 - \log \sigma^2_{\vecx, i}
    \bigr)}_{\equiv \mathcal{D}
    _{\mathrm{KL}}^{(i)} \ge 0}
\end{equation}

In the Bayesian sense, $\mathcal{N}(0, I)$ is to be interpreted as a prior for the latent space distribution, and we penalize the network for deviating from this prior~\cite{doersch_tutorial_2021}. In practice,
this regularizes the latent space learned by the model by pushing it to only use as much of the latent space as necessary to perform reconstruction well, as well as pushing the latent variables to be statistically independent. The relative strength of 
the KL divergence is tuned by the regularization hyperparameter
$\lambda.$
The appropriate scale of $\lambda$ is set by the input and latent dimensionality as well as the overall scale of the inputs, but in practice must be tuned by hand (see \App{SM:VAE_framework}).
During training, we find it useful to anneal the regularization strength by starting training with $\lambda = 0$ and slowly
increasing it until reaching
the final (reported) value~\cite{bowman_generating_2016}. 
We perform this optimization by standard minibatch training using the ADAM optimizer~\cite{kingma_adam_2017} (see \App{SM:ML_architecture}).

\section{Results \& Discussion}

\subsection{Training results}

\begin{figure}[t]
\centering
\includegraphics[width=0.9\columnwidth]{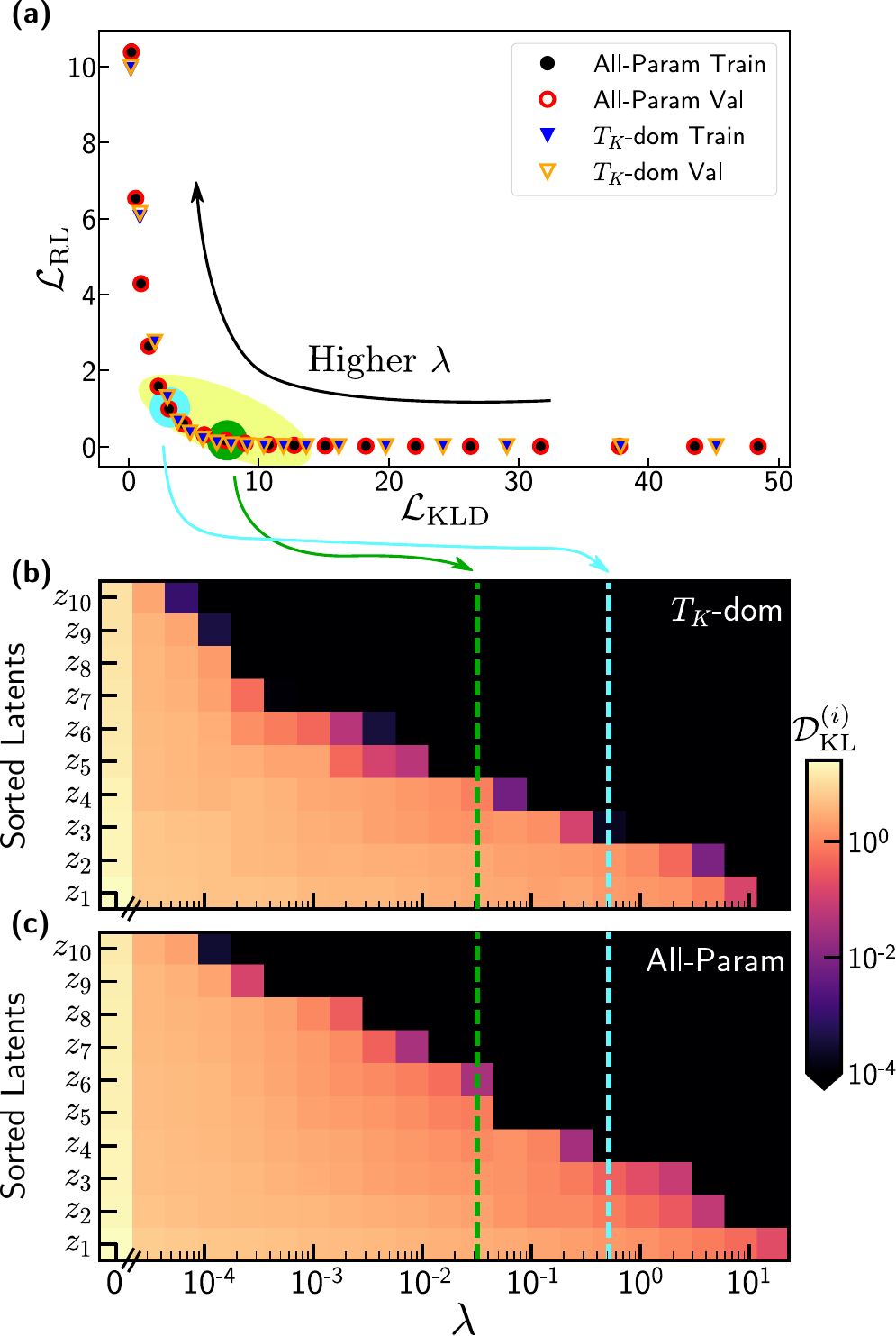}
\caption{
(a) Final $\mathcal{L}_{\mathrm{RL}}$ reconstruction and $\mathcal{L}_{\mathrm{KLD}}$ regularization losses of $L=10$ VAEs as the regularization strength $\lambda$ increases. The highlighted yellow region is a visual aid showing the ``critical region'' of models with good $\mathcal{L}_{\mathrm{RL}}$-$\mathcal{L}_\mathrm{KLD}$ tradeoffs. The green circle marks the center of this critical region identifiable as an inflection point in log-log space
(see \App{SM:Training_details}),
while the cyan circle marks models with stronger regularization which are presented in this work. 
(b,c) Average KL loss $\mathcal{D}_{\mathrm{KL}}^{(i)}$
contributed by each individual latent variable $z_i$ as $\lambda$
increases, for models trained on (b) the $T_K$-dominated and (c) the full-parameter datasets. Latent variables are ordered by descending
$\mathcal{D}_{\mathrm{KL}}^{(i)}$ loss.
Values shown are from the models with the fewest active $z_i$ across three training runs at each $\lambda$.
}
\label{fig:train_results}
\end{figure}

To search for interpretable models that capture meaningful aspects of the dataset, we first train a collection of $L = 10$ VAEs at various final regularization strengths $\lambda.$ For each value of $\lambda$, we train three models \footnote{We observe that the variance in final losses between trained models is negligible for the purposes of constructing the loss curves in Fig.~\ref{fig:train_results}(a). However, due to multiple local minima, at some $\lambda$ there is occasionally $\pm 1$ active latent in the final model.} and measure the final reconstruction and KLD losses, averaged across the dataset. 
Plotting these against each other in Fig.~\ref{fig:train_results}(a)
demonstrates the fundamental reconstruction-regularization trade-off. Ideally, we want to minimize both losses to obtain a model which captures the dataset with a simply structured, low-dimensional latent space.

An interesting feature of this trade-off is a
pronounced corner region (highlighted in yellow)
where $\rloss + \klloss$ is minimal.
Moving towards large $\lambda$ only improves $\klloss$ modestly, but incurs a large sacrifice in reconstruction performance.
Since $\mathcal{D}_\mathrm{KL}$ measures the amount of information encoded into the latent space relative to the unit Gaussian prior~\cite{doersch_tutorial_2021}, we interpret
the region marked in yellow as where the ``critical'' amount of
information is present in the model. Beyond this point, the information capacity of the model dips below what is necessary to capture the trends contained in the dataset. In Sec.~\ref{SM:VAE_framework}, we show that this critical region is clearly identifiable as an inflection point in log-log space. This is understandable as a local optimum in $(d \rloss / d \klloss) / (\rloss/\klloss)$,
which is a scale-invariant measurement of the tradeoff magnitude.

We can gain further insight into these models by observing how the total $\mathcal{D}_{\mathrm{KL}}$ is spread across the latent variables for each $\lambda,$ as shown in \Fig{fig:train_results}(b). Here, and for the rest of the paper, we sort the latent variables
$z_i$ by their respective average $\mathcal{D}_{\mathrm{KL}}^{(i)},$ in decreasing order.
Increasing the regularization strength has the effect of entirely deactivating some of the latent neurons, resulting in them simply predicting the prior $z_i \sim \mathcal{N}(0, 1)$
irrespective of the input $\vecx$, thus
achieving a nearly-zero KL loss
$\mathcal{D}_{\mathrm{KL}}^{(i)}.$ 

As visualized in \Fig{fig:train_results}(b), the number of active
neurons in latent space is rather sharply defined
for finite $\lambda$.
We find that at the onset of the ``critical'' $\lambda$ region, the number of active neurons nearly matches
the number of free physical parameters describing the relevant training dataset.
Three physical parameters $\{U, \Gamma, \epsilon_d\}$ characterize the $T_K$-dominant dataset, and we observe that the number of active neurons in the relevant trained model transitions from $4\to 3 \to 2$. Similarly, the full-parameter dataset has two additional underlying parameters $\{B, T\}$, and the trained model transitions from $6\to 5 \to 4$ active $z_i$. We find however, that these discovered latents do not correlate well one-to-one with the bare Hamiltonian parameters, reflecting that bare parameters are not necessarily the best way of characterizing the dataset.

A second desired effect of the KL regularization is to push each dimension of the latent space to correspond a ``disentangled'' feature~\cite{higgins_early_2016} which carries a distinct physical meaning. Ideally, this results in each latent variable $z_i$ learning a statistically independent feature, i.e. $\mathrm{Cov}(z_i, z_j) = \delta_{ij}\mathrm{Var}(z_i)$. In \App{SM:Training_details}, we define a heuristic metric of success towards this goal as a normalized sum of off-diagonal elements of the latent covariance matrix, and show that statistical independence indeed improves as $\lambda$ is increased. To obtain simple interpretable models which are well-disentangled and low-dimensional while also capable of reconstructing spectra with reasonable accuracy, we train beyond the critical $\lambda$ at $\lambda \approx 0.5$ (cyan markers in \Fig{fig:train_results})
where only $2, 3$ active $z_i$ remain for the $T_K$-dominated and full-parameter sets, respectively. Since these models have fewer active $z_i$ than the number of free physical parameters we know to be present, we are effectively forcing the models to learn high-level combinations of these parameters, corresponding to the most important physical variations in the dataset. (See \App{SM:LatentTraversals} for views of lower-regularization models).

\subsection{Interpretation of the latent spaces}

We first investigate VAE models trained to reconstruct spectral functions from the $T_K$-dominated regime,
$T_K \gg T, |B|$. To analyze the latent space structure post-training, we disable the random sampling by setting
$\vecz = \boldsymbol{\mu}_\vecx$.
Hence, when discussing distributions below, we are referring the distribution of $\boldsymbol{\mu}_\vecx$ in the latent space across all $\vecx$ in the training dataset. For the $T_K$-dominated dataset, we find that well-trained models at $\lambda \approx 0.5$ have all but two bottleneck neurons deactivated [cf. \Fig{fig:train_results}(b)].

\begin{figure}[]
    \centering
    \includegraphics[width=\columnwidth]{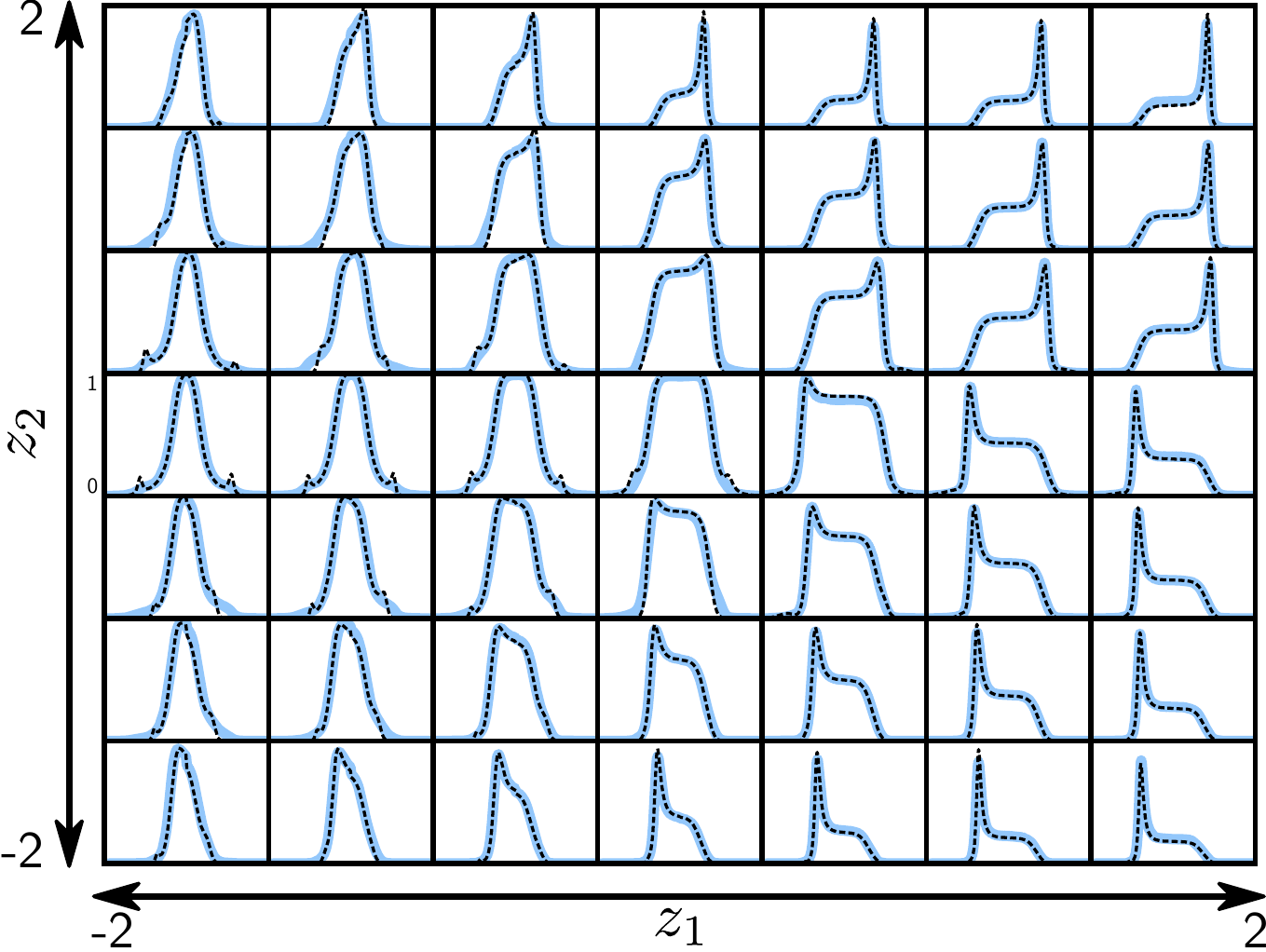}
    \caption{Thick blue curves show reconstructed spectral functions corresponding to different points in the two-dimensional latent space learned by a $\lambda = 0.5$ VAE on the $T_K$-dominated dataset. Thin dashed black curves show real spectral functions in the dataset whose compressed latent representations are closest to the sampled latent space point. Curves are plotted on the log-linear frequency grid (see \App{SM:Training_details}) as they are fed into the VAE.
}
\label{fig:lowBlowT_sweep}
\end{figure}

To understand the nature of this two-dimensional active latent
space, we scan the values of the two active latents and analyze how the reconstructed spectral functions evolve.
Examples are shown in \Fig{fig:lowBlowT_sweep}, where we sweep the two active neurons in the range $z_i \in [-2, 2]$ corresponding to two standard deviations of the KL prior.
We also provide the respective ground-truth input NRG spectra which encodes the closest $\vecz$-value. We observe that the reconstructed spectra smoothly evolve as we move around this space,
reconstructing plausible intensity profiles. However,
due to the reconstruction vs. KL loss trade-off, this model
misses the fine details of the small Hubbard-like side peaks at large energies for large negative $z_1$.

\Fig{fig:lowBlowT_sweep} shows that increasing $z_1$  corresponds to a broadening and flattening of the central Kondo peak in the spectral function. Physically, we infer that this feature relates to the Kondo temperature $T_K$ which sets the overall energy
scale associated with this peak. Meanwhile, $z_2$ learned 
the physical particle-hole asymmetry of the SIAM
Hamiltonian. For any $(z_1, z_2)$,
the transformation $z_2 \rightarrow -z_2$ roughly results in the reflection of the decoded spectral function around $\omega = 0$.

\begin{figure}[]
    \centering
    \includegraphics[width=\columnwidth]{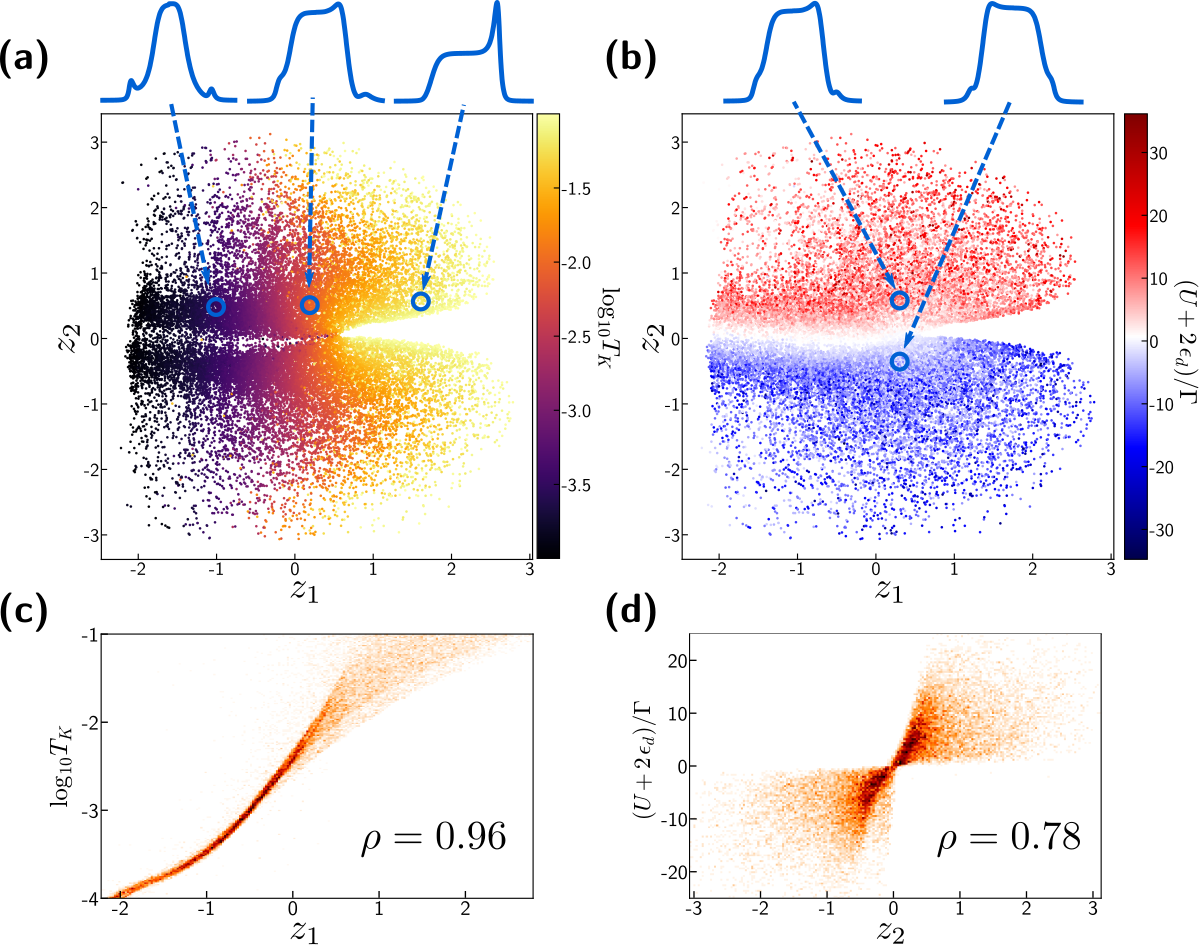}
    \caption{Latent space learned by a $\lambda = 0.5$ VAE
    on the $T_K$-dominated
    dataset, colored by (a) $\log_{10}(T_K)$
    and (b) $(U + 2\epsilon_d)/\Gamma$. The blue open circles mark the points in latent space corresponding to the shown spectral functions. In (c) and (d), correlations between the principal components and $\log T_K, (U+2\epsilon_d)/\Gamma$ are plotted, whose strengths are quantified by Pearson correlation coefficients
    $|\rho| \le 1$ which are measures of linear correlation.
}\label{fig:lowB_latent}
\end{figure}

These correspondences are summarized in \Fig{fig:lowB_latent} where we
plot the distribution of the training dataset in latent space and color the points according to relevant physical parameters. We find a remarkably high correlation between the first latent dimension and $\log T_K$
[\Fig{fig:lowB_latent}(a,c)]. Furthermore, in \Fig{fig:lowB_latent}(b,d) we see that $z_2$ can be associated with
the ratio $(U + 2\epsilon_d)/\Gamma$ which characterizes the particle-hole asymmetry in the SIAM. These strong correlations indicate that the VAE has successfully learned the dominant physical features characterizing the dataset.

\begin{figure}[t!]
    \centering
    \includegraphics[width=\columnwidth]{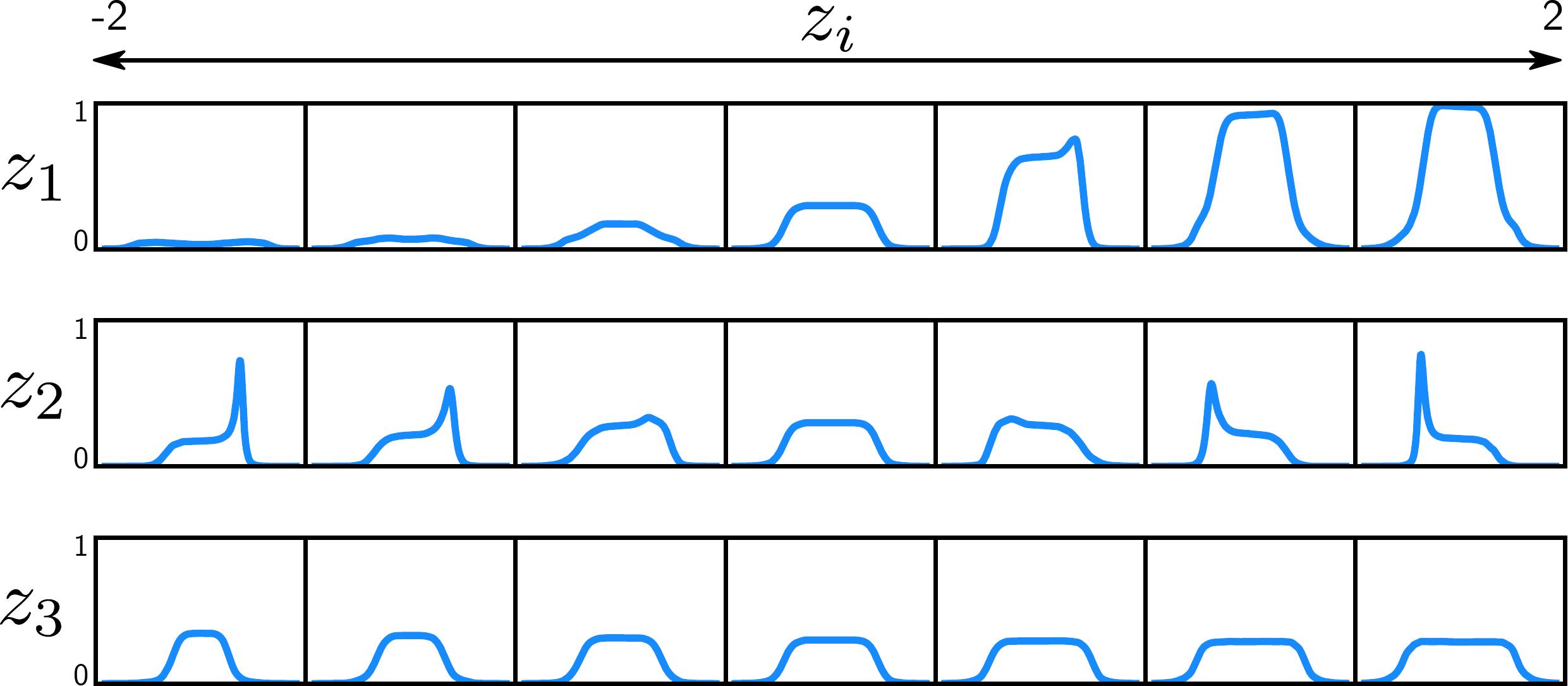}
    \caption{Reconstructed spectral functions from a $\lambda = 0.3$ VAE trained on the full five-parameter dataset. Each row corresponds to a line scan in latent space where all $z_i$ are held at $0$ except for one.
    }
    \label{fig:fiveparam_sweep}
\end{figure}

\begin{figure}[]
\centering
\includegraphics[width=0.9\columnwidth]{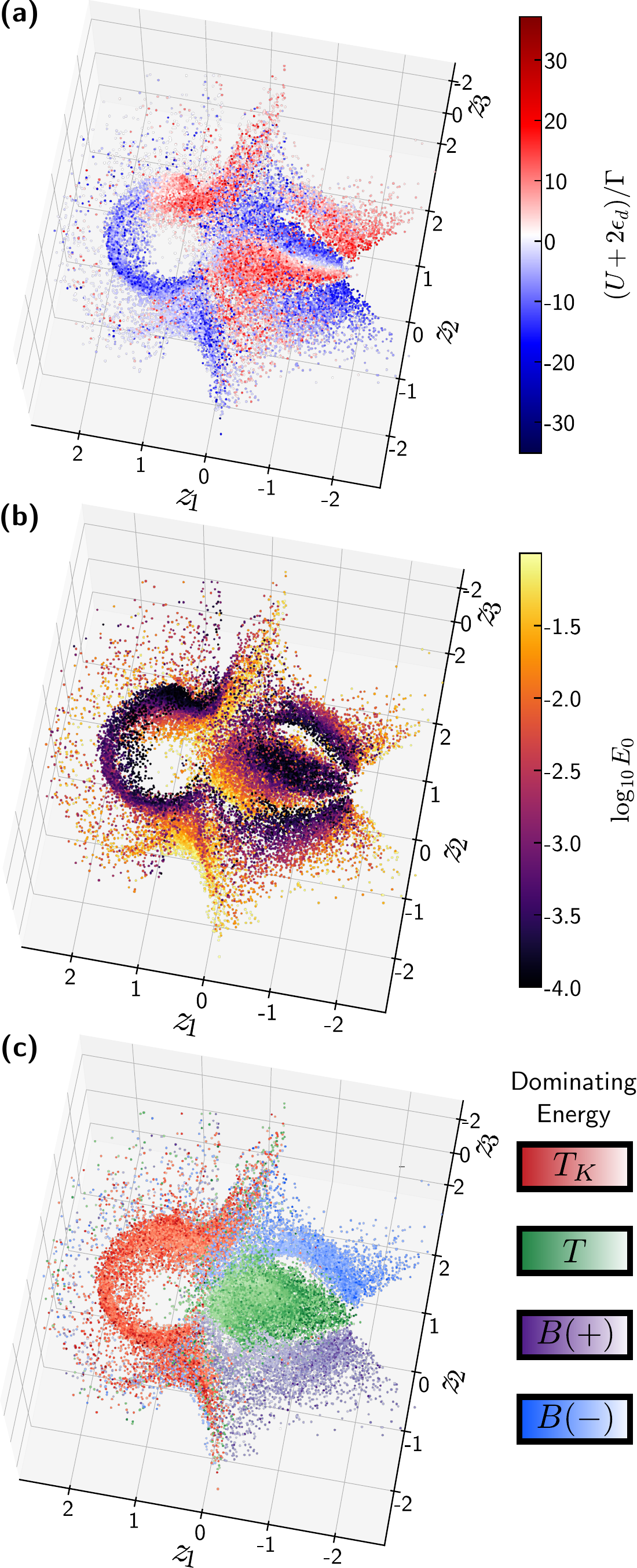}
\caption{%
    Latent space learned by the VAE on the full five-parameter
    dataset using $\lambda = 0.3$. The data in (a-c) is precisely the same,
    except that the color coding differs based
    on the chosen physical parameter:
    (a) $(U + 2\epsilon_d)/\Gamma$;
    (b) $\log_{10}\mathrm{E_0}
      = \log_{10}\max(T_K, T, |B|)$;
    (c) dominant energy scale out of $\{T_K, T, B\}$. Within each
    cluster, points are darker if the relative factor to the next-highest energy scale is larger.
    Supplementary views from different angles are provided
    in \Fig{fig:supp_3d_view_angles}.
}
\label{fig:fiveparam_latent}
\end{figure}

We now investigate VAEs trained on the full five-parameter dataset, which shares the SPE control equally between $\lvert B\rvert, T, \text{and } T_K$.
We again choose to examine models from the $\lambda \approx 0.5$ cyan region of Fig.~\ref{fig:train_results}, where we see only three latents remain active. Similar to Figs.~\ref{fig:lowBlowT_sweep},\ref{fig:lowB_latent}, we simultaneously examine both direct generative scans of the learned latent space (Fig.~\ref{fig:fiveparam_sweep}), as well as the dataset distribution in the full three-dimensional latent space (Fig.~\ref{fig:fiveparam_latent}). Interestingly, we now find that rather than one globally continuous manifold, our models tend to break the dataset into multiple distinct clusters which only separately form continuous spaces. This decomposition can be directly connected to certain physical properties in the dataset. In Fig.~\ref{fig:fiveparam_latent}(c), we show how the learned latent space decomposes into four disconnected clusters, each of which is characterized by a different physical parameter
that dominates the smallest physical energy $E_0$. Additional views of these data are available in \App{SM:AlternateViews}.

When all $z_i=0$ (center column of Fig.~\ref{fig:fiveparam_sweep}),
the height of the spectral function
is reduced (indicative of finite temperature $T$), the spectrum
is left-right (particle-hole) symmetric,
and shows no Hubbard side peaks (indicative of
larger $\Gamma/U$).
Now when varying $z_1$, this changes the height
of the spectral function while keeping it largely
symmetric. Therefore $z_1$ controls the underlying
temperature, which is also reinforced by observing that $z_1\lesssim 0$
is in a $T$-dominated regime,
as seen by the green data in \Fig{fig:fiveparam_latent}
and \Fig{fig:supp_3d_view_angles}.
On the other hand, large positive $z_1$ switches
towards the $T_K$-dominated low-$T$ regime
(red data in \Fig{fig:fiveparam_latent}).
Additionally, we find that $z_3$ controls the width
of the spectral function's central peak. As such, it
is directly connected to the absolute energy scale describing each spectral function, as shown in \Fig{fig:fiveparam_latent}(b) and \Fig{fig:supp_3d_view_angles}.

Meanwhile, $z_2$ controls much of the asymmetry of the spectral
function around $\omega = 0$. At small $B$, $z_2$,
characterizes the particle-hole asymmetry
$(U + 2\epsilon_d)/\Gamma$ [see \Fig{fig:fiveparam_latent}(a)].
Magnetic fields also break the inversion symmetry of the
spin-resolved spectral data. Hence $B$ competes with the effect
of particle-hole asymmetry controlled by
$(U + 2\epsilon_d)/\Gamma$. For large $\pm B$
therefore the latent space switches to a $B$-dominated
domain (see blue data in \Fig{fig:fiveparam_latent}(c))
which goes hand in hand with a sign change in $z_2$.

Due to the rather strong regularization, the examined VAE does
not fully capture all of the features of the dataset. In particular, this examined model focuses on the central Kondo peak
features, but tends to miss the sharp Hubbard-like side peaks 
appearing in the high-$T$ and high-$B$ regions. In Fig.~\ref{fig:supp_latent_traversals} of the Appendix, we examine VAEs trained with weaker regularization that do successfully capture these side peaks, but are more difficult to interpret.

To gain confidence that our VAEs learn meaningful representations, an important question
is whether the structures of the data manifolds discovered by these separately-trained VAEs are consistent with each other. The $T_K$-dominated dataset by its construction lies entirely within the red high-$T_K$ ``bulb'' of the full-parameter dataset at large $z_1$ in Fig.~\ref{fig:fiveparam_latent}, seen to be a roughly two-dimensional manifold. Do continuous paths within the two-dimensional latent space of the $T_K$-dominated model correspond to continuous paths on this bulb?

\begin{figure}[]
\centering
\includegraphics[width=0.9\columnwidth]{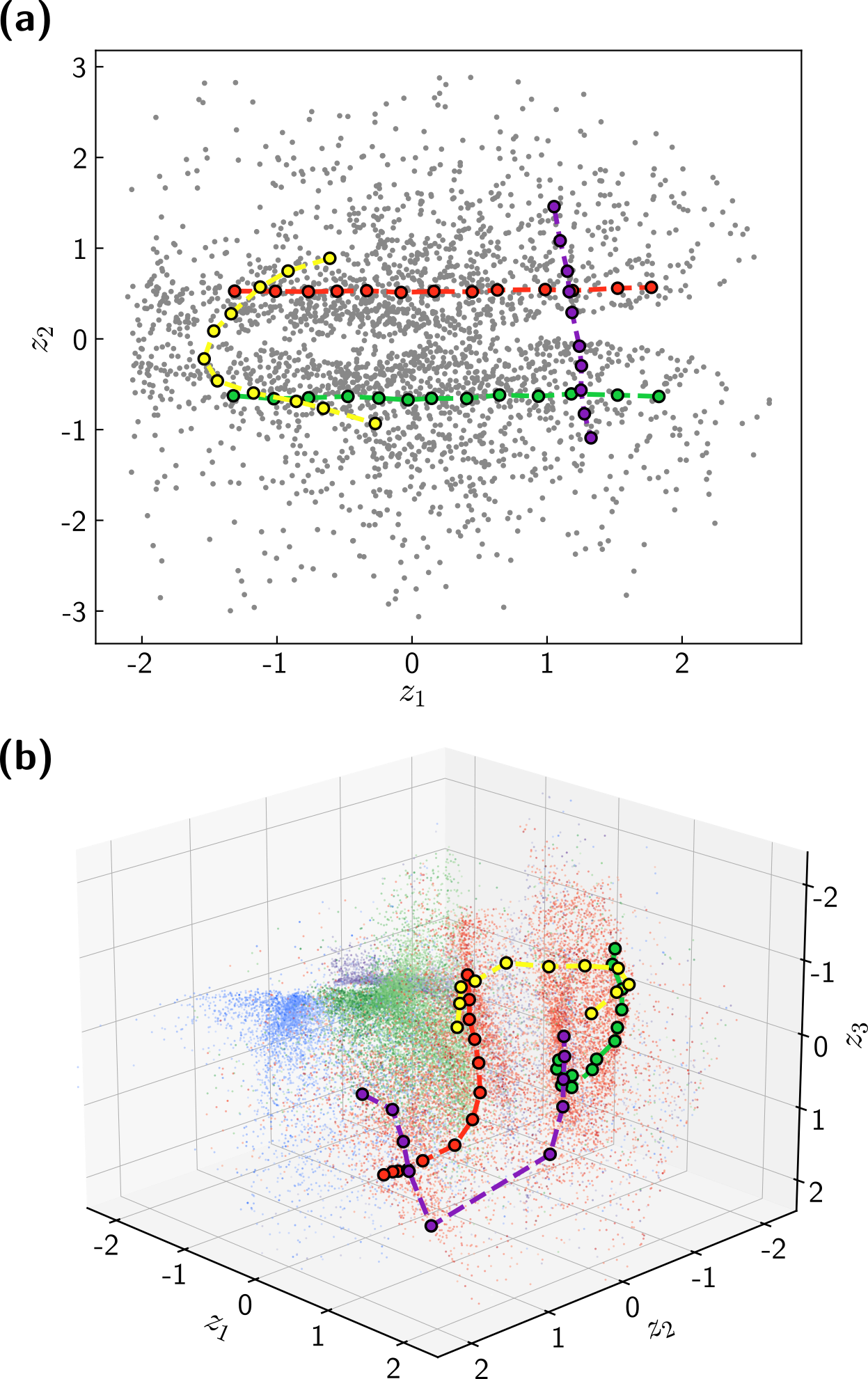}
\caption{%
   (a)
   Chosen paths in the
   latent space of the VAE trained on the $T_K$-dominated
   dataset. Each highlighted point corresponds to a single spectral function in
   the dataset. 
   (b)
   The same spectral
   functions mapped into the latent space learned by the VAE
   trained on the full-parameter dataset, colored as in \Fig{fig:fiveparam_latent}(c). Supplementary view angles are
    provided in \Fig{fig:supp_3d_view_angles}.
}
\label{fig:latent_mapping}
\end{figure}

To answer this, we start by choosing
several paths of points in the latent space of the $T_K$-dominated model, as shown in \Fig{fig:latent_mapping}(a).
The red and green paths are chosen to be at roughly constant and opposite values of the particle-hole asymmetry $z_2$, extending primarily along $z_1 \sim \log T_K$. The yellow and purple paths are chosen to span the two particle-hole asymmetric lobes, with the yellow path tracing through a seemingly connected region while the purple path crosses an apparent void. By feeding these same spectra into the encoder of the full-parameter VAE, we can retrace these paths in its learned three-dimensional activate latent space.

As shown in \Fig{fig:latent_mapping}(b),
we indeed find that the mapped path
demonstrates consistency in the sense of continuity
and qualitative topology between the two models.
The two-dimensional $T_K$-dominated latent space is continuously ``wrapped''
into the red frontal bulb of the full-parameter latent space.
Interestingly, the purple path maintains its single discontinuity
in the extended space, occurring exactly when the path jumps
between the two particle-asymmetric lobes.
This analysis further confirms that these learned latent spaces are all meaningfully connected to the same underlying physics, regardless of which subset of the data is seen by the model.

\subsection{Symbolic regression to extract physical descriptors}

Much of our previous analysis depended on having the physical foresight to determine explicit physical features which could correspond to various dimensions in latent space. For use in machine learning-aided \textit{discovery}, we require some means to automatically extract these features given a trained VAE. We propose a potential route towards this goal using \textit{symbolic regression} (SR)~\cite{schmidt_distilling_2009, bongard_automated_2007}. We assume knowledge of a set of physical parameters $\vecp$
associated with the generation of each spectral function (and hence each latent 
vector $\mathbf{z}$). In our case $\vecp$ will be the Hamiltonian parameters of Eq.~\eqref{Eq-SIAM}, while in experimental settings these may be, e.\,g., the temperature and applied pressure. Using SR, we can then search through the space of analytic expressions of $\vecp$ to find those which best describe each learned latent $z_i$. Intuitively, we hope that the information the VAE has chosen to keep in the bottleneck will inform us of effective combinations of $\vecp$ which correspond to meaningful physical quantities. While SR has been previously investigated as a way to understand the \textit{transformation} learned by the network \cite{CranmerNEURIPS2020}, our goal here is different -- the network has never seen $\vecp$, and we are using SR to inform us how the critical bottleneck information relates to the original generative factors $\vecp$.

\begin{figure}[]
    \centering
    \includegraphics[width=0.9\columnwidth]{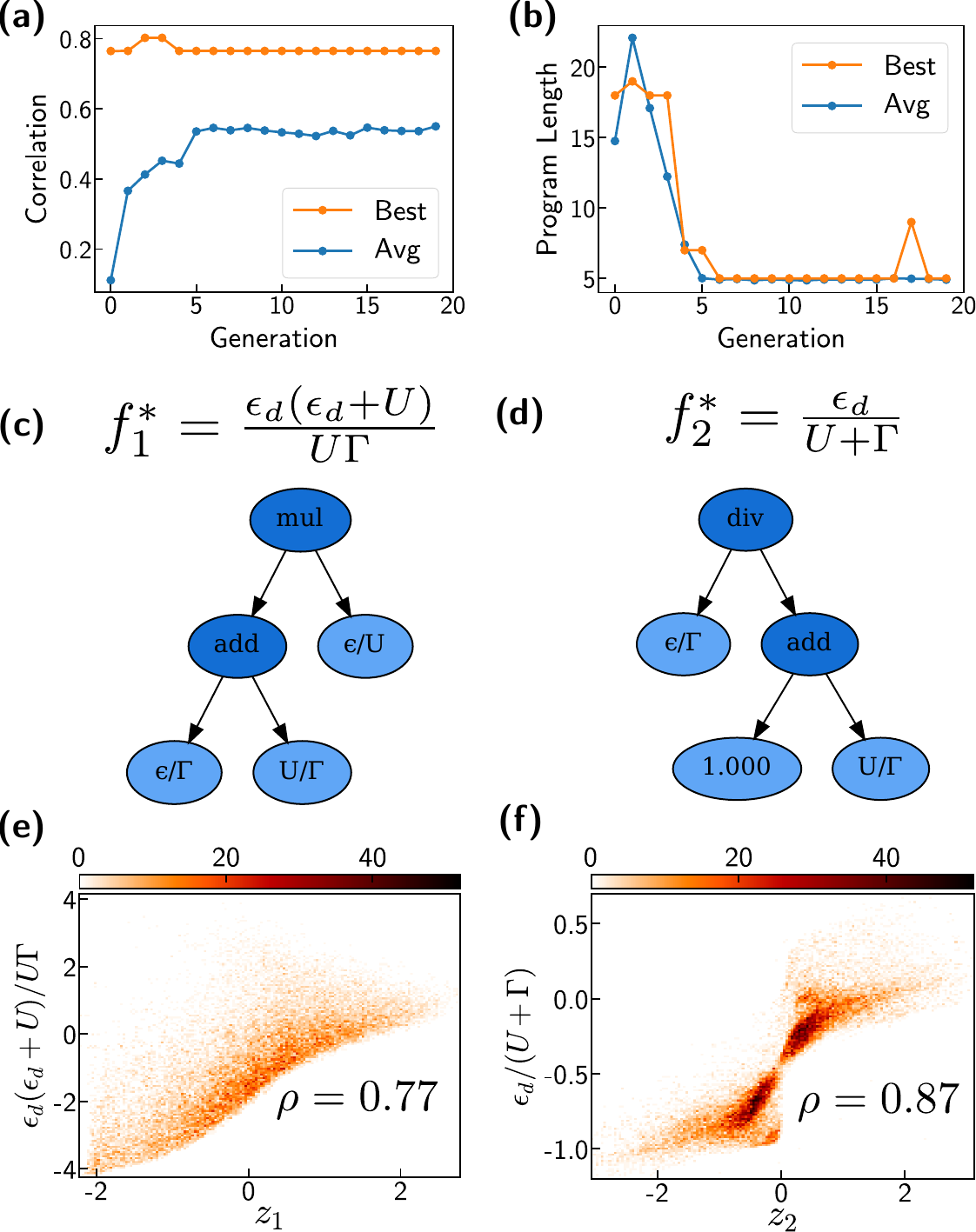}
    \caption{Results of applying symbolic regression to learn functions describing each latent variable of the VAE trained on the $T_K$-dominated dataset (Fig.~\ref{fig:lowB_latent}). (a) Symbolic regression population fitness and (b) program length as the algorithm learns $f_1^\ast$. (c,d) Obtained symbolic syntax trees and equivalent functions which maximize correlation with the active latents $z_i$. (e,f) Scatterplots showing $f^\ast_i$ vs. $z_i$ for each latent dimension, analogous to Fig.~\ref{fig:lowB_latent}(c, d). Colorbars show the density of data at each point.
}
    \label{fig:symb_reg}
\end{figure}

In SR, functions are represented as syntax trees, where
operations appear as interior nodes, and parameters $p_i$ appear
as leaves [see Fig.~\ref{fig:symb_reg}(c,d) for examples]. We define the class of representable functions $\mathcal{S}$ by specifying a list of unary and binary functions which are allowed to appear in these trees. In this work, we limit this list to include the standard binary arithmetic operations $\{+, -, \times, /\}$ as well as unary negation and inversion. Then, we phrase the objective as finding the symbolic function $f_i(\vecp)$ in this class which maximizes the Pearson correlation coefficient with an individual latent $z_i$:
\begin{equation} \label{eq:symb_reg}
    f^\ast_i(\vecp) = \arg\max_{f_i \in \mathcal{S}} \frac{\mathbb{E}_{\vecp \sim D}[(f_i(\vecp) - \expval{f_i})(z_i(\vecp) - \expval{z_i})]}{\sigma_{f_i} \sigma_{z_i}},
\end{equation}
where expectation values indicate means with respect to the dataset and $\sigma$'s are standard deviations of the subscripted variables. We note that this objective function is invariant under an overall scale factor or constant shift of either the $f$'s or $z$'s. To bias the optimization process towards simple expressions, an additional penalty is introduced of the form $\gamma l$ where $\gamma$ is a hyperparameter and $l$ is the number of nodes in the syntax tree.

One symbolic function $f_i^\ast$ is regressed per active latent dimension $z_i.$ We utilize the \texttt{gplearn}
library~\cite{stephens_gplearn_2015}, which performs this search using genetic algorithm techniques~\cite{pal_genetic_1996}. At the beginning of the regression, a large number $N_{\mathrm{pop}}$ of trial symbolic functions are added to a population by constructing random syntax trees up to a given finite depth. At each stage, the worst-performing functions are discarded and new functions are generated by randomly modifying the best-performing functions using biologically inspired \textit{mutation} and \textit{crossover} operations. Mutation operations randomly replace nodes with alternate operations or parameters, while crossover operations mix subtrees between functions. Iterating these processes improves the general performance of the entire population until reaching some plateau. While this approach is generally sensitive to the specification of various hyperparameters controlling mutation/crossover probabilities, we find that other than increasing the population size to $5000$ and the program length penalty to $\gamma=0.01$, the default package parameters perform well for our purposes.

We apply this approach to extract explicit functions describing
the latent dimensions of VAEs trained on the $T_K$-dominated
dataset in \Fig{fig:lowB_latent}. Since $\abs{B},T \ll T_K$
in this dataset, the available Hamiltonian parameters which may appear in our regressed functions are $\{U, \Gamma, \epsilon_d\}$. We find it useful to enforce unitless expressions by setting the available leaf nodes in the regression to be the unitless ratios $\vecp = \{U/\Gamma, \epsilon_d/U, \epsilon_d/\Gamma\}$. We regress two functions $f_1^\ast(\vecp), f_2^\ast(\vecp)$ to obtain symbolic expressions which correlate well with the two active latent variables $z_1, z_2$, resulting in Fig.~\ref{fig:symb_reg}.

Remarkably, we consistently find the function automatically discovered as describing $z_1$ to be $f_1^\ast = \frac{\epsilon_d(\epsilon_d + U)}{U\Gamma}$ (syntax tree in Fig.~\ref{fig:symb_reg}(c)), which is precisely
$\log T_K$ [\Eq{eq:TK}] up to a logarithmic correction of $\log(U\Gamma)$ inexpressible in terms of our chosen operator set, and a constant coefficient/shift which leaves the correlation invariant. While the symbolic regression does not directly inform us that it has learned the $\log$ of a relevant quantity, this can be inferred from the fact that the data has been presented to the VAE on a logarithmic energy grid. The correlation between this optimized function $f_1^\ast$ and $z_1$ was found to be $\rho = 0.77$ (Fig.~\ref{fig:symb_reg}(e)) while the correlation between the full $\log T_K$ and $z_1$ is $\rho = 0.96$, indicating that these logarithmic corrections explain a small but notable fraction of the correlation. We find that expanding the operator set to include $\log$ and using bare input parameters tends to recover these key logarithmic factors, but also generates a handful of extraneous terms. We anticipate that careful tuning or advanced approaches producing Pareto frontiers \cite{deb_nsgaii_2002} would successfully filter through this excess, though are beyond the scope of this work.

The obtained function which correlates best with $z_2$ was consistently found to be $f_2^\ast = \epsilon_d / (U + \Gamma)$, which differs from our ``knowledgable'' guess for particle-hole asymmetry of $(U + 2\epsilon_d) / \Gamma$. In fact, $f_2^\ast$ is found to have a greater correlation with $z_2$ than $(U+2\epsilon_d) / \Gamma$ does, showing that this is not a failure of the regression. This instead can be explained from the observation that the latent space is not very well-structured at large $|z_2|$, as seen from the large spread in
Fig.~\ref{fig:lowB_latent}(b,d). While the VAE can accurately perform the correct $z_2$ ordering of spectral functions which slightly deviate from the particle-hole symmetric point, it seems to not focus as much on getting this ordering correct between very asymmetric curves. Motivated by this observation, we find that if the symbolic regression is restricted to regress only the high-density region of points with $|z_2| < 0.5$, it then does reliably recover the ``ideal'' expression of $(U + 2\epsilon_d)/\Gamma$.

\section{Discussion}

The results of this work demonstrate a promising avenue towards machine-assisted physical discovery. Given an unlabeled set of data generated through some physical process, our work shows how methods from unsupervised machine learning can automatically extract physically-meaningful structure. VAEs accomplish this by discovering a small set of disentangled features parameterizing a low-dimensional manifold which can be used to reconstruct the data faithfully. However, alternate dimensionality reduction techniques can also stand in this place (see \App{SM:PCA}). Symbolic regression then allows us to discover an (approximate) analytic parameterization of this manifold in terms of a set of known physical parameters, which here was found to near-exactly reproduce a known complex physical descriptor in the single-impurity Anderson model. While here we demonstrate these approaches on spectral function data, we believe this to be a powerful idea applicable for general physical data with appropriate modification of the architecture.

As presented in this manuscript, our procedure uses simple ``off-the-shelf'' techniques from machine learning and therefore is easily transferable to a wide set of problems. While this was already sufficient to discover significant and meaningful structure in the current dataset, we envision several improvements that can be made. Interesting possible modifications include specializing the VAE architecture to better capture the target data, or changing the VAE prior itself which may be beneficial in datasets where the underlying descriptors are not necessarily Gaussian-distributed~\cite{mathieu_disentangling_2019}. A potential example use-case is in phase identification~\cite{Wetzel2017Phys.Rev.E}, where the latent variables may fall into disjoint clusters depending on the phase attributed to the input. Additionally, more complex~\cite{deb_nsgaii_2002, petersen2021deep} or physically-motivated~\cite{Udrescueaay2631, reinbold_robust_2021} approaches to symbolic regression may produce more robust and meaningful symbolic expressions to describe the latent space. This may include the ability to produce a full Pareto frontier \cite{deb_nsgaii_2002} of increasingly complex but accurate formulae. Finally, a remaining task is a thorough investigation of the minimal data needed to extract physical insights, which is relevant in lower-data experimental settings. We leave these developments to future work.

\section*{Data \& Code Availability}

The data used in this study has been made available at Ref.~\cite{ZenodoDataset}. The code used to train the VAEs and perform the symbolic regression in this study has been made available at \href{https://github.com/ColeMiles/SpectralVAE}{https://github.com/ColeMiles/SpectralVAE}.

\begin{acknowledgements}
CM and MRC acknowledge the following support: This material is based upon work supported by the U.S. Department of Energy, Office of Science, Office of Advanced Scientific Computing Research, Department of Energy Computational Science Graduate Fellowship under Award Numbers DE-FG02-97ER25308 and DE-SC0020347.
EJS and RK were supported by the U.S. Department of Energy, Office of Science, Basic Energy Sciences as a part of the Computational Materials Science Program. AW was supported by the U.S. Department of Energy, Office of Basic Energy Sciences. This research used resources of the Center for Functional Nanomaterials, which is a U.S. DOE Office of Science Facility, and the Scientific Data and Computing Center, a component of the Computational Science Initiative, at Brookhaven National Laboratory under Contract No. DE-SC0012704. KB acknowledges  support from the Center of Materials Theory as a part of the  Computational Materials Science (CMS) program, funded by the U.S. Department of Energy, Office of Basic Energy Sciences.

Disclaimer: This report was prepared as an account of work sponsored by an agency of the United States Government. Neither the United States Government nor any agency thereof, nor any of their employees, makes any warranty, express or implied, or assumes any legal liability or responsibility for the accuracy, completeness, or usefulness of any information, apparatus, product, or process disclosed, or represents that its use would not infringe privately owned rights. Reference herein to any specific commercial product, process, or service by trade name, trademark, manufacturer, or otherwise does not necessarily constitute or imply its endorsement, recommendation, or favoring by the United States Government or any agency thereof. The views and opinions of authors expressed herein do not necessarily state or reflect those of the United States Government or any agency thereof.
\end{acknowledgements}


\appendix

\renewcommand{\thefigure}{A\arabic{figure}}

\section{Training Details} \label{SM:Training_details}

\subsection{Training data \label{SM:coarse_graining}}

\begin{figure}[bh!]
    \centering
    \includegraphics[width=\columnwidth]{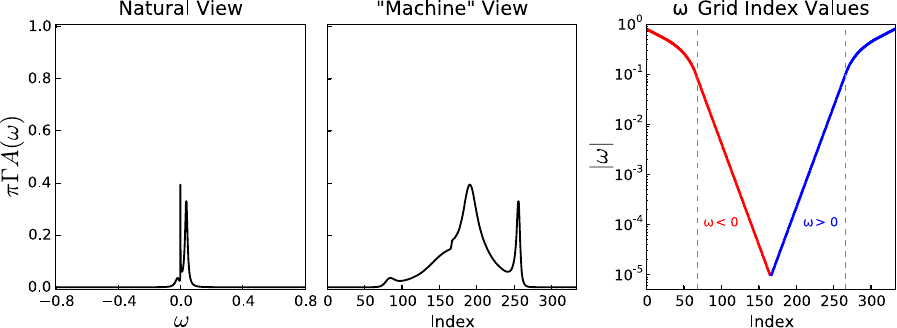}
    \caption{An example of the coarse-grain mapping from $A\left(\omega\right)$ to $\vecx$. The left panel shows the original spectral function on a linear $\omega$ axis. The middle panel showcases how the same spectrum distorts under the linear-logarithmic $\omega_i$ grid, forming the input $\vecx$ to the VAE. The right panel shows the frequency values for the 333 coordinates with the exception of the single bridge point at $\omega\sim10^{-11}$. 
    }
    \label{fig:omega_grid}
\end{figure}

Each trial in the dataset is represented by a 
discretized subset of coordinate pairs
$\bigl(\omega_i, \pi\Gamma A(\omega_i) \bigr)$.
We only examine the spectral weight between $\omega \in [-0.8, 0.8]$ to avoid band-edge artifacts caused by the sharp 
band edge of the otherwise featureless hybridization functions.  Within this ``window,'' we select $i=1,\ldots, 333$ frequency points on a mixed linear-logarithmic grid which are the same for all trials. Between $\lvert\omega\rvert \in [0.1, 0.8]$, the $\omega$ grid is spaced linearly, at $\delta\omega_i \sim 0.01$ intervals; between $\lvert\omega\rvert \in[10^{-5},10^{-1}]$, the grid is at even intervals in log$_{10}$ space. A single point at $\omega\approx 0$ bridges the gap between negative and positive frequencies. This procedure of coarse-graining $A\left(\omega\right) \to \vecx$ provides sufficient
resolution
for our spectral features of interest
at a very manageable modest grid
size. For more details, see Fig.~\ref{fig:omega_grid} and Appendix
in \cite{Sturm21}.

\subsection{VAE framework}
\label{SM:VAE_framework}

Here, we briefly connect our abridged presentation of the VAE in the main text to the ``standard'' probabilistic presentation in the literature, roughly following Ref.~\cite{doersch_tutorial_2021}.
In the context of generative modeling, one assumes that the data $D$ is sampled from some unknown probability distribution, $P(\vecx)$, which we would like to infer. As a latent variable model, VAEs factorize this distribution as 
\begin{equation}
    P(\vecx) = \int d\vecz \: P(\vecx | \vecz)P(\vecz),\label{eq:supp_vae_factor}
\end{equation}
where $\vecz\sim\mathcal{N}(0, I)$ are normally-distributed latent variables controlling the generation of $\vecx$. Both $P(\vecz | \vecx)$ and $P(\vecx | \vecz)$ are modeled as neural networks (the encoder and decoder, respectively), which we denote in this section as $\mathrm{Enc}(\vecz | \vecx)$ and $\mathrm{Dec}(\vecx | \vecz)$. We denote the distribution over $\vecx$ predicted by a model in training as $\tilde{P}(\vecx) = \int d\vecz \: \mathrm{Dec}(\vecx | \vecz) P(\vecz)$, which we aim to drive towards the true $P(\vecx)$.

We would like to minimize the negative log-likelihood of the data $D$, $-\mathbb{E}_{\vecx \sim D}[\log \tilde{P}(\vecx)]$ to optimize this variational distribution.
However, a full evaluation of this requires an expensive marginalization over $P(\vecz)$ as in Eq.~\eqref{eq:supp_vae_factor}. Instead, simultaneously learning $P(\vecz | \vecx)$ through the encoder allows us to minimize the negative of the easily-calculable ``evidence lower-bound'' (ELBO) which lower-bounds the log-likelihood \cite{kingma_auto-encoding_2014}:

\begin{multline}
    \mathcal{L} = -\mathbb{E}_{\vecx \sim D}\Big[ \mathbb{E}_{\vecz \sim \mathrm{Enc}}\big[\log \mathrm{Dec}(\vecx | \vecz) \\ - \mathcal{D}_{\mathrm{KL}}[\mathrm{Enc}(\vecz | \vecx) || \mathcal{N}(0, I)] \big]\Big]. \label{eq:supp_elbo}
\end{multline}
To arrive at the loss presented in the main text, the encoder is taken to model a normal distribution in latent space, $\mathrm{Enc}(\vecz | \vecx) = \mathcal{N}(\boldsymbol{\mu}_{\vecx}, \mathrm{diag}(\boldsymbol{\sigma}_{\vecx}))$, and the decoder is interpreted as modeling a Gaussian distribution centered at the prediction $\tilde{\vecx}$ as $\mathrm{Dec}(\vecx | \vecz) = \mathcal{N}(\tilde{\vecx}, \lambda I/2)$, with $\lambda$ a hyperparameter. Inserting these into Eq.~\eqref{eq:supp_elbo} brings the loss into the form of Eq.~\eqref{eq:loss:tot} shown in the main text.

Within this perspective, the hyperparameter $\lambda$ may be roughly interpreted as how accurate we can expect the decoder's predictions to be, given access to $\vecz$. An alternate perspective (taken from Ref.~\cite{higgins2016beta}, where a similar parameter $\beta$ is employed) is that $\lambda$ regularizes the model to emphasize making the latents normally-distributed relative to making accurate reconstructions. To maintain a fixed reconstruction/regularization balance, $\lambda$ should be linearly related to the input dimension and inversely related to the chosen latent dimension \cite{higgins_early_2016}. At the same time $\lambda$ must be increased with the average scale of $\vecx$. In practice, this means that some amount of hyperparameter scanning is required, as we exhibit in Fig.~\ref{fig:train_results}(a).

\begin{figure}[h!]
    \centering
    \includegraphics[width=\columnwidth]{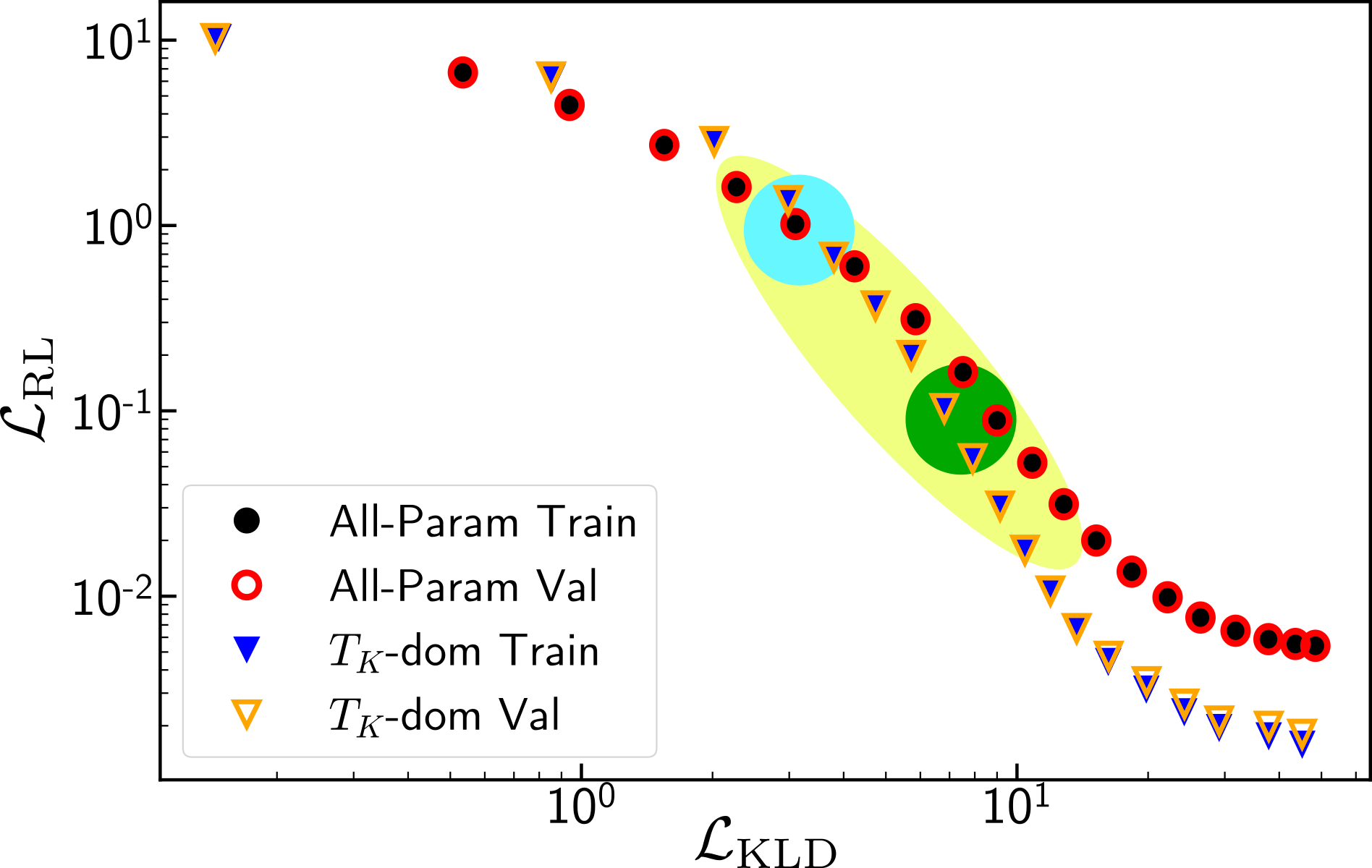}
    \caption{The same performance measurements and region highlights as in Fig.~\ref{fig:train_results}(a), but on a log-log scale. Here, we can identify the ``critical region'' as a pronounced inflection in the $\rloss-\klloss$ curve.}
    \label{fig:supp_log_log_perf}
\end{figure}

By scanning multiple models trained at various $\lambda$ values, we find that a reasonable method for identifying critical VAE regions is to plot $\rloss$ versus $\klloss$ as in Fig.~\ref{fig:train_results}(a), but on a log-log scale as in Fig.~\ref{fig:supp_log_log_perf}. Notably, the shape of this curve is independent of an overall scaling of the data, which simply shifts the curve in log-log space. We note that the critical region appears as a clearly defined inflection point, which we can identify as a local optimum in the first derivative. To select our final models shown in the main text, we train multiple randomly-initialized VAEs with the same $\lambda$, and choose the model with the fewest active latent neurons and the simplest structure in latent space.

\subsection{Machine Learning Implementation}
\label{SM:ML_architecture}

We construct our neural networks using the PyTorch~\cite{paszke_pytorch_2019} deep learning library, and train them using standard minibatch training with the ADAM~\cite{kingma_adam_2017} optimizer for a total of $E=2000$ epochs. Every epoch, the training dataset is randomly partitioned into ``minibatches'' with \texttt{batch\_size = 1024} spectral functions. Model parameters are updated once per batch using gradients backpropagated from the error on the samples within the minibatch. Step sizes are adaptively computed within the ADAM algorithm, with an upper bound set by the learning rate (\texttt{lr}) hyperparameter. This \texttt{lr}, initially set to $\texttt{lr}=0.001$, is modified each epoch according to a cosine annealing schedule \cite{loshchilov_sgdr_2016} as implemented by Pytorch's \texttt{CosineAnnealingLR}. Between each fully connected layer of our network, we employ a batch normalization \cite{ioffe_batch_2015} layer to promote gradient backflow. Our code is made public at \href{https://github.com/ColeMiles/SpectralVAE}{https://github.com/ColeMiles/SpectralVAE}.

\begin{figure*}
    \centering
    \includegraphics[width=1.7\columnwidth]{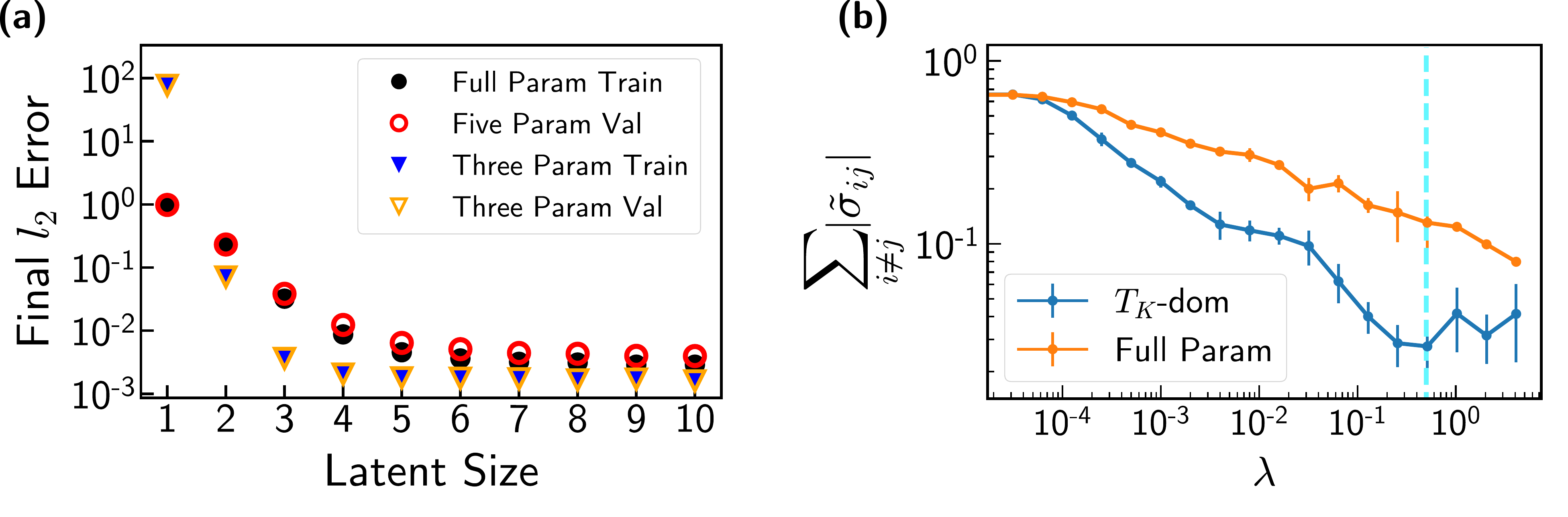}
    \caption{Additional performance measurements. (a) Final reconstruction performance of unregularized $(\lambda = 0)$ VAEs as the latent dimension $L$ is varied. (b) Measurements of our disentangling metric upon increasing $\lambda$,
    showing that the $\mathcal{D}_{\mathrm{KL}}$ loss effectively drives learned features to be statistically independent. The cyan dashed line marks $\lambda = 0.5$.}
    \label{fig:supp_performance}
\end{figure*}

As mentioned in the main text, we anneal the regularization strength $\lambda$ by ramping up from $0$ to the final value across the first fraction of the training epochs. This is heuristically found to improve the quality of trained VAEs \cite{bowman_generating_2016}, due to avoiding local minima where the model solely focuses on minimizing $\mathcal{D}_{\mathrm{KL}}$. The exact form of this ramping is not found to significantly affect results as long as it is done slowly enough, and we employ a ramp of the form $\lambda\tanh(4t/E)$ with $t$ the current epoch number.

In Fig.~\ref{fig:supp_performance}(a), we show final reconstruction performance results for unregularized ($\lambda = 0$) VAEs as $L$ is varied. When the latent dimension size $L$ is large, we note a significant plateau in reconstruction performance. This plateau is already firmly established by $L = 10$, motivating our choice to set this as the latent size of our VAEs. We also note a ``critical'' latent $L$, smaller than which reconstruction performance rapidly worsens. Interestingly, this is roughly $3$ and $5$ for the $T_K$-dominated and full-parameter datasets respectively, corresponding to the number of free physical parameters in each case. This provides an alternate method to extract the critical dimension from that shown in the main text.

Once we turn on the $\mathcal{D}_{\mathrm{KL}}$ regularization loss (i.e. $\lambda \neq 0$), 
this drives the learned latent features $z_i$ to
become statistically independent. In the VAE literature, statistically independent features are referred to as ``disentangled'' \cite{higgins_early_2016, mathieu_disentangling_2019}. We define a heuristic metric of success towards this goal as the normalized sum of the off-diagonal elements of the covariance matrix. With $\sigma_{ij} = \mathbb{E}[(z_i - \langle z_i\rangle)(z_j - \langle z_j\rangle)]$ being the standard covariance, we define normalized elements as:

\begin{equation}
    \tilde{\sigma}_{ij} = \frac{\sigma_{ij}}{\sum_{i'j'}|\sigma_{i'j'}|},
\end{equation}
and define a ``disentangling'' metric as $\sum_{i\ne j} \tilde{\sigma}_{ij}$, which becomes zero when all $z_i$'s are completely independent. In Fig.~\ref{fig:supp_performance}(b), we show that this metric rapidly decreases with increasing $\lambda$, demonstrating how the regularization pushes the latent $z_i$'s towards statistical independence.

\section{Alternate Views of Latent Space} \label{SM:AlternateViews}

\begin{figure*}
    \centering
    \includegraphics[width=1.8\columnwidth]{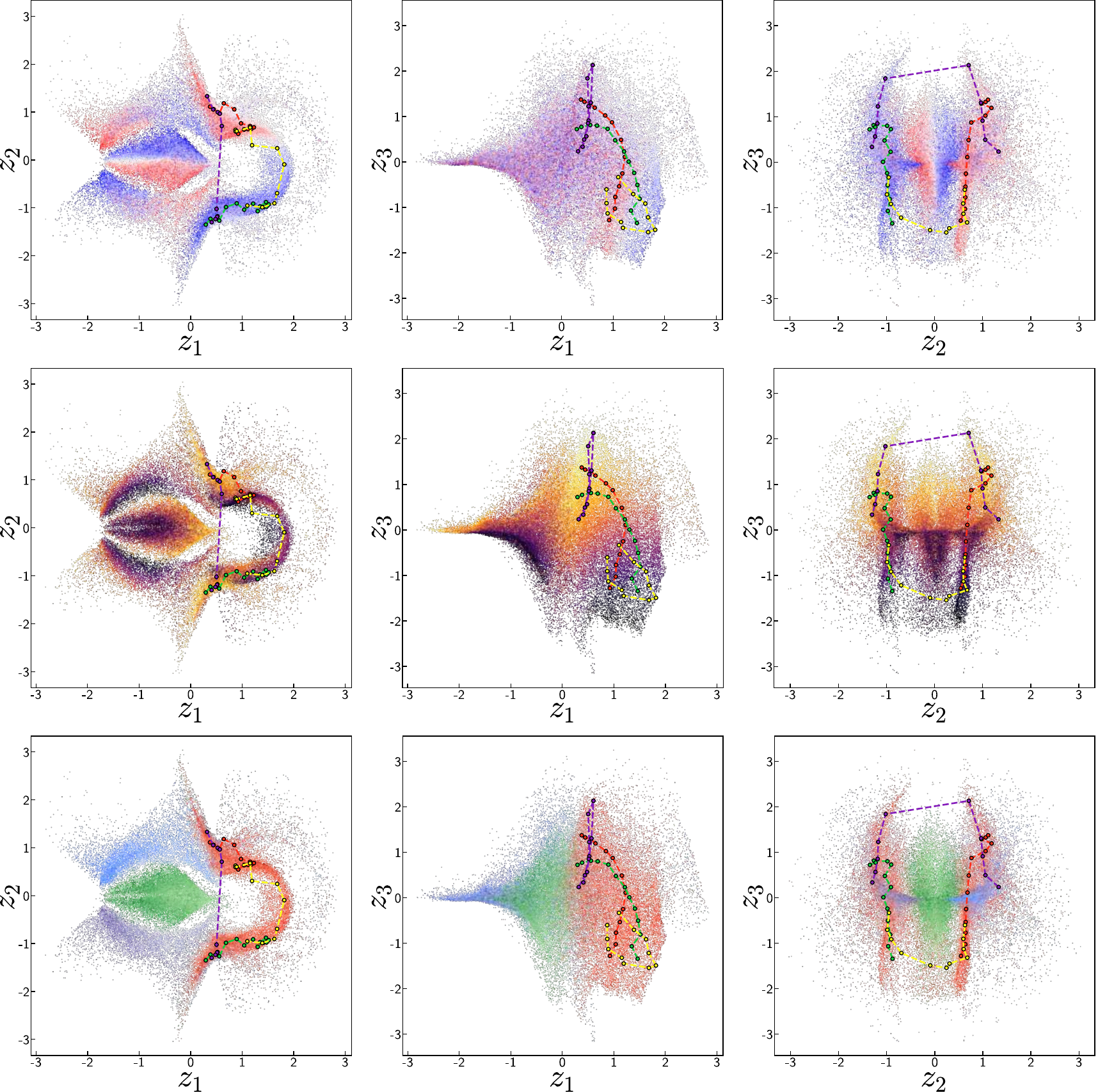}
    \caption{Additional view angles of each plane in the three-dimensional latent space learned by the VAE on the full five-parameter dataset in the main text, with the same colorings as in Fig.~\ref{fig:fiveparam_latent}.}
    \label{fig:supp_3d_view_angles}
\end{figure*}

Here we offer the reader additional viewpoints into the VAE-learned latent spaces characterizing the full-parameter dataset. In particular, \Fig{fig:supp_3d_view_angles} shows multiple plane projections of the VAE presented in the main text (\Fig{fig:fiveparam_latent}). Each column shows a different plane projection onto two of the active latents, while each row colors the points according to $(U+2\epsilon_d)/\Gamma, \log_{10} E_0$, or the dominant energy scale $\{T_K, |B|, T\}$, with each colormap as in \Fig{fig:fiveparam_latent}. A particular feature which is more evident here is the strong correlation between the latent $z_2$ and the smallest physical energy scale $\log E_0$ (second row).

\begin{figure}[h!]
    \centering
    \includegraphics[width=0.97\columnwidth]{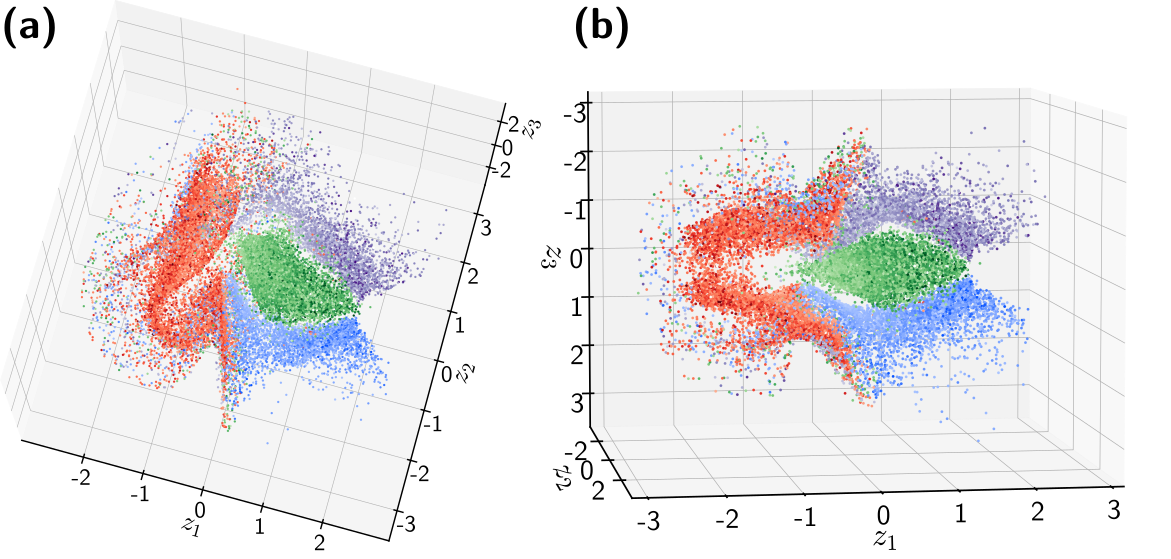}
    \caption{Latent spaces of two additional randomly initialized, independently trained $\lambda = 0.3$ VAEs on the full-parameter dataset, with coloring as in Fig.~\ref{fig:fiveparam_latent}(c). Both VAEs displayed here have a higher total loss than the one presented in the main text.}
    \label{fig:supp_more_vaes}
\end{figure}

Next, in \Fig{fig:supp_more_vaes} we show the learned latent spaces of two additional randomly initialized, independently trained VAEs at the same value of $\lambda = 0.3$. Both of the models presented have a higher total loss than the VAE presented in the main text. However, they can be seen to have qualitatively similar (but lower-quality) structure in latent space. On rare occasions the learned structure will ``twist'' into an extra dimension, but these are easily identifiable by a large $\mathcal{D}_{\mathrm{KL}}$ loss. In practice, due to the presence of multiple local minima in the loss function it is necessary to train multiple models and analyze the best-performing one.

\section{Comparison to linear dimensionality reduction} \label{SM:PCA}

Since we do not make use of the generative aspects of VAEs in this work, their application here can be understood as an example in the broader class of ``dimensionality reduction'' or ``representation learning'' \cite{bengio_representation_2013} techniques. All methods falling into this category attempt to find a low-dimensional parameterization of a collection of high-dimensional data. Due to their disentangling property, we find VAEs are particularly well suited for discovering independent, physically meaningful descriptors, but other dimensionality reduction techniques may also fit into our pipeline. To demonstrate this, we compare our results from the main text to the results of a linear dimensionality reduction approach provided by principal component analysis (PCA).

Principal component analysis is a simple, parameter-free technique for discovering the directions in a feature space which vary the most within a dataset. Thinking of each input spectral function as a $333$-dimensional vector $\vecx$, PCA is done by simply diagonalizing the covariance matrix $C_{ij} = \langle (x_i - \langle x_i\rangle)(x_j - \langle x_j \rangle)\rangle$, with the expectations taken over the dataset. The resulting eigenvectors can be thought of as a rotated orthogonal basis of this vector space ordered such that the first basis vector points along the direction of greatest variance in the dataset (with its corresponding eigenvalue the variance), and then decreasing in variance for each subsequent vector. A dimensionality reduction can then be achieved by projecting the original data onto the first few of these principal component vectors.

\begin{figure*}[h]
    \centering
    \includegraphics[width=0.9\textwidth]{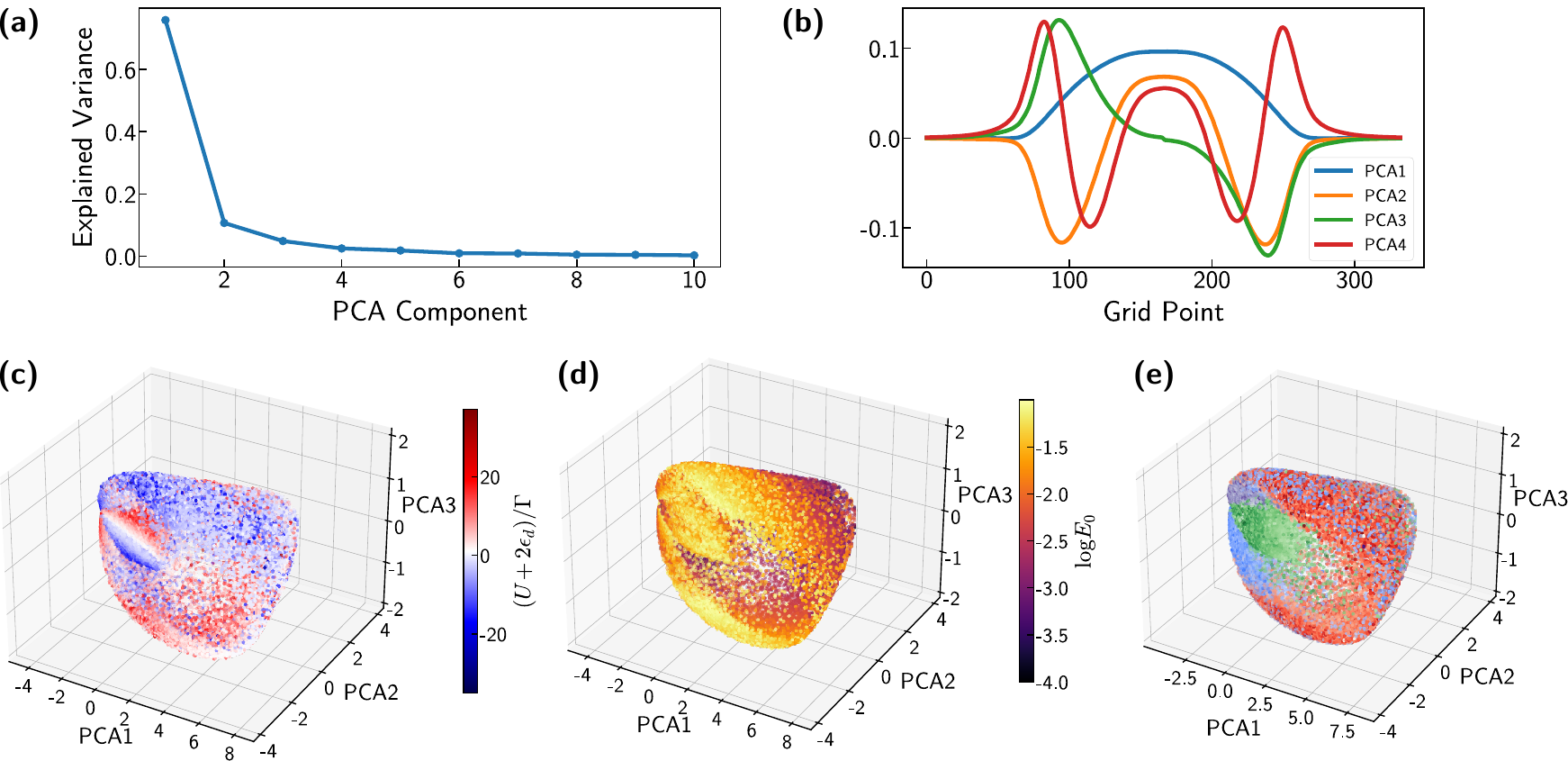}
    \caption{Results of applying principal component analysis (PCA) directly to the spectral functions in the full five-parameter dataset. (a) Explained variance, defined as the fraction of the variance along each PCA axis compared to the sum of all variances, for each discovered principal component. (b) The first four principal component vectors. (c-e) Scatter plots of projections of the dataset onto the first three principal components, colored respectively by the particle-hole asymmetry,
    $\log_{10} E_0$, and the dominant energy scale [with color coding as in \Fig{fig:fiveparam_latent}(c)].
    }
    \label{fig:supp_pca_curves}
\end{figure*}

We show the results of applying PCA to the full-parameter dataset in \Fig{fig:supp_pca_curves}. In Fig.~\ref{fig:supp_pca_curves}(a), we show that a great fraction of the total variance of the data is captured by only a few principal components, four of which we plot in Fig.~\ref{fig:supp_pca_curves}(b). We can see that the first of these components (blue) measures the broad central spectral weight, the second and fourth (orange, red) characterize the width of the central peak, and the third (green) characterizes some measure of asymmetry. We note that our dataset as a whole is very nearly symmetric about the central frequency -- for every spectra in the dataset there exists another spectra which is approximately its mirror about the center. As a result, all of the principal components are either symmetric or antisymmetric about the center.

\begin{figure*}[h]
    \centering
    \includegraphics[width=0.9\textwidth]{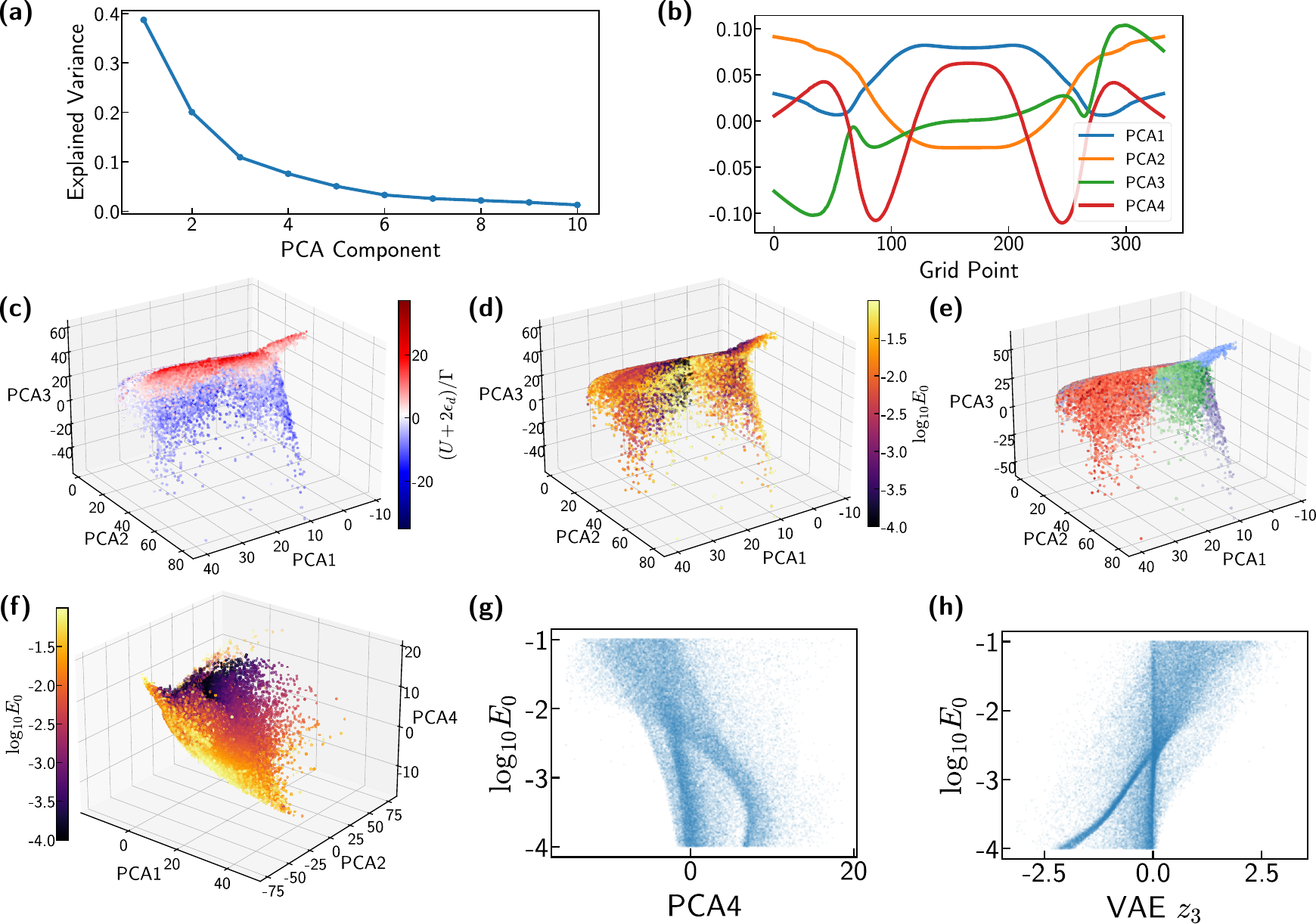}
    \caption{Like \Fig{fig:supp_pca_curves}, but with a frequency-dependent normalization performed before PCA. (a--e) are as in \Fig{fig:supp_pca_curves}. (f) An alternate 3D projection, showing that PCA4 roughly captures $\log_{10}E_0$. (g,h) Scatterplots showing correlation between $\log_{10} E_0$ and either PCA4 or the VAE's latent $z_3$ (introduced in the main text), respectively.
    }
    \label{fig:supp_pca_curves_normed}
\end{figure*}

While these features are somewhat similar to those discovered by the VAE, the reduced space is obtained only from linear projections onto these curves. We find that this is sufficient to somewhat capture the particle-hole asymmetry and overall energy scale (Fig.~\ref{fig:supp_pca_curves}(c,d)), but the separate energy scales $T_K, |B|, T$ are folded together in a somewhat complex way (Fig.~\ref{fig:supp_pca_curves_normed}). Evidently, some nonlinear transformations are needed to properly unfold this structure into one where all physically meaningful descriptors are naturally aligned with the new dimensions. A virtue of the VAE process is that we can arrive at a \textit{minimal} set of such features, even capturing more complex features are shown in the following section.

A typical method for applying PCA is to perform some amount of feature normalization as a preprocessing step. One standard scheme is to normalize the spectra such that each frequency is independently normalized to zero-mean and unit-variance across the dataset, i.e. $\tilde{x}_i \gets (x_i - \langle x_i\rangle)/\sigma_{x_i}$. This normalization approach effectively emphasizes subtle details in the tails of $\vecx$, putting them on the same scale as other changes. In Fig.~\ref{fig:supp_pca_curves}, we show the result of applying PCA to the full five-parameter dataset after this normalization has first been performed. We see in Fig.~\ref{fig:supp_pca_curves}(b) that again the first and third components measure the broad central spectral weight and asymmetry respectively, while the the second and fourth (orange, red) now characterize the amount of weight in the tails compared to the center.

Surprisingly, this new normalization process has made various energy scales separate out far more cleanly. In particular, the various dominating energy scales are now visible as distinct lobes in \Fig{fig:supp_pca_curves_normed}(e). Additionally, the smallest physical energy $E_0$ now appears correlated with PCA4 as shown in \Fig{fig:supp_pca_curves_normed}(f,g), though not quite as sharply as the correlation with the VAE-discovered latent $z_3$ (\Fig{fig:supp_pca_curves_normed}(h)). We remark that although the VAE did not have the benefit of this human-inspired normalization scheme, it was nonetheless able to discover a set of features that directly correlates with known physics.

\section{Latent Traversals of Weaker-Regularization Models} \label{SM:LatentTraversals}

\begin{figure*}[h!]
    \centering
    \includegraphics[width=0.9\textwidth]{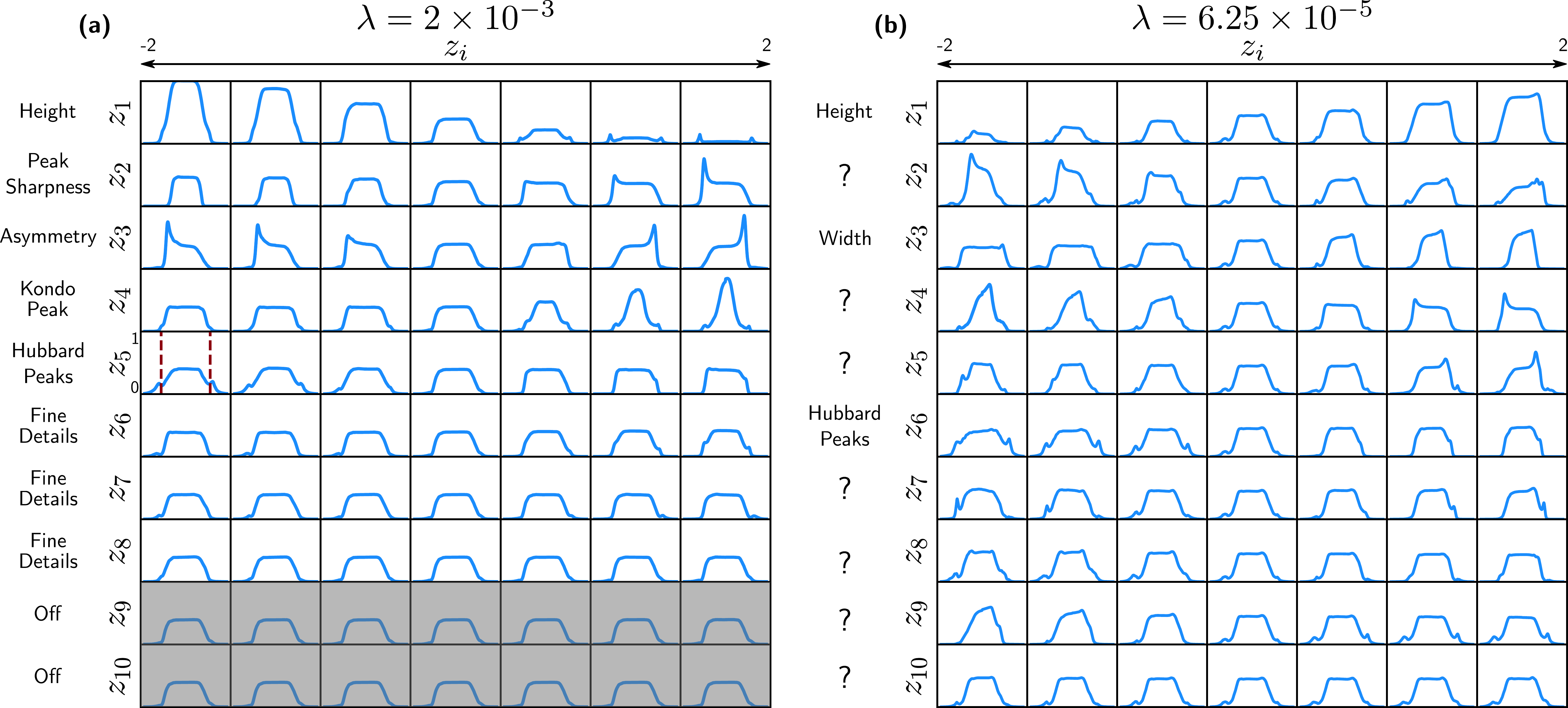}
    \caption{Line scans along each latent variable $z_i$ of VAEs trained on the full five-parameter dataset for smaller regularization strengths than the $\lambda \approx 0.5$ examined in the main text
    having (a) $\lambda=0.002$ and (b) $\lambda=6.25\times 10^{-5}$. The latent variables $z_i$ have been sorted according to decreasing KL loss, as in the main text. Red dashed lines mark the log-to-linear frequency grid crossover points described in \ref{SM:coarse_graining}.
    Both models can be seen to recover finer peak details
    as compared to \Fig{fig:fiveparam_sweep}, though are not quite as well disentangled.
    }
    \label{fig:supp_latent_traversals}
\end{figure*}

In the main text, for simplicity of presentation we examined VAEs trained with a rather strong regularization of $\lambda \approx 0.5$ which were found to smoothly capture high-level information about our dataset. However, as a consequence of this strong regularization our models miss
smaller features such as details of the Hubbard side peaks in the spectral functions. Indeed, it is a well-known phenomena that VAE reconstructions tend to appear ``blurred'' compared to the original input \cite{higgins2016beta}.
In \Fig{fig:supp_latent_traversals}, we examine VAEs trained with weaker regularization strengths by scanning each latent
dimension $z_i \in [-2, 2]$ while keeping all others at zero. We find that at intermediate regularization we can capture additional meaningful features in the dataset beyond those presented in the main text, demonstrating that we are not limited by the learning capacity of our VAE.

For $\lambda = 0.002$ in \Fig{fig:supp_latent_traversals}(a),
we find that two latent variables are completely inactive, and two additional variables ($z_7, z_8$) barely affect the reconstruction. For the remaining active neurons, we can still (partially) assign meaningful descriptors even at this weaker regularization. Additionally, we can tell from our disentangling metric (Fig.~\ref{fig:supp_performance}(b)) that these variables are more correlated with each other than at $\lambda = 0.5$. In particular, these additional active latents appear to capture fine details about the development of the Hubbard side peaks in the spectra.

At even smaller $\lambda = 6.25\times10^{-5}$
in \Fig{fig:supp_latent_traversals}(b), we see
that all latent variables seem to affect the reconstruction in at least a minor way. Additionally, we see that it is difficult to separate the latent variables into 
particular physical effects on the reconstruction, as multiple variables can be seen to produce similar changes on the resulting spectral function.

\clearpage

\bibliography{bib}

\end{document}